\documentclass[a4paper]{article}

\usepackage[margin=2.5cm]{geometry} 
\usepackage{setspace}
\usepackage[utf8]{inputenc}
\usepackage[T1]{fontenc}

\usepackage{amsmath, amsthm} 
\usepackage{amssymb}
\usepackage{mathrsfs}
\usepackage{mathtools}
\usepackage{optidef} 
\mathtoolsset{showonlyrefs=true}

\usepackage{titlesec}
\titleformat*{\section}{\large\bfseries}
\titleformat*{\subsection}{\normalsize\bfseries}

\titleformat{\paragraph}[runin]
  {\normalfont\slshape}
  {\theparagraph}
  {1em}
  {}
  [. ]

\DeclareMathOperator{\Var}{Var}

\usepackage{graphicx}
\usepackage{caption}
\usepackage{subcaption}
\usepackage{booktabs}
\usepackage{threeparttable}
\usepackage{siunitx}
\usepackage{multirow}
\usepackage{tabularx}
\usepackage{etoolbox}
\robustify\bfseries

\usepackage[round,authoryear]{natbib}

\usepackage[colorlinks=true,allcolors=blue]{hyperref}

\usepackage{lineno}

\usepackage{authblk}

\theoremstyle{plain}

\theoremstyle{definition}

\title{\large \textbf{
The Macroeconomic Cost of Network Impedance 
}}
\author[a]{Satoshi Nakano\thanks{Email: nakano@n-fukushi.ac.jp}}
\author[b]{Kazuhiko Nishimura\thanks{Email: nishimura@lets.chukyo-u.ac.jp 
}}
\affil[a]{Faculty of Economics, Nihon Fukushi University, Tokai 477-0031 Japan}
\affil[b]{Institute of Economics, Chukyo University, Nagoya 466-8666 Japan}

\date{\small \today} 

\begin{document}

\maketitle


\begin{abstract}
We quantify the macroeconomic and welfare costs of input-output network rigidity under empirical input complementarity ($\sigma_i < 1$). Evaluating expected aggregate productivity conditional on avoiding negative equilibrium singularities, we show that symmetric Gaussian microeconomic shocks induce a negative drift in aggregate total factor productivity---termed \textit{mean sinking}---governed by structural \textit{network impedance}. Calibrating a $174$-sector general equilibrium model to U.S.\ input-output accounts and embedding the resulting shock distribution into a global Real Business Cycle model, we find that macroeconomic persistence and capital accumulation amplify this drag into a permanent consumption-equivalent welfare loss of roughly $-7\%$ relative to a Cobb-Douglas benchmark. Finally, lengthening intermediate supply chains magnifies this welfare burden under input complementarity while leaving expected welfare invariant under Cobb-Douglas technology.
\end{abstract}
\section{Introduction}
Understanding how microeconomic disturbances in individual industries propagate through supply chains to shape aggregate fluctuations and social welfare is a central question in modern macroeconomics. In a frictionless economy characterized by Cobb-Douglas production technologies and unitary elasticities of substitution, Hulten's (1978) theorem implies that idiosyncratic sectoral productivity shocks aggregate log-linearly according to their Domar weights. Under this classical benchmark, symmetric zero-mean microeconomic shocks cancel out in expectation, leaving the mean path of aggregate total factor productivity (TFP) undisturbed, while second-order consumption fluctuations around a fixed trend carry negligible welfare costs \citep{Lucas1987}.

A growing body of literature, however, challenges this log-linear paradigm by demonstrating that production networks transmit and distort microeconomic shocks nonlinearly. \citet{Acemoglu2017} show that within Cobb-Douglas input-output networks, macroeconomic tail risks---such as severe left-tail contractions in aggregate output---can emerge from non-Gaussian, fat-tailed distributions of underlying microeconomic shocks. Moving beyond unitary elasticity, \citet{BaqaeeFarhi2019} establish that when sectoral production functions exhibit constant elasticity of substitution (CES) different from unity, nonlinearities proportional to $(1 - \sigma)$ alter the tail thickness of aggregate output and shift expected aggregate productivity away from zero even when microeconomic innovations are perfectly symmetric. Drawing on empirical evidence that elasticities of substitution across intermediate inputs in the U.S.\ economy lie predominantly below unity ($\sigma < 1$), their second-order approximation reveals that input complementarity imparts a downward bias to average aggregate productivity. More recently, \citet{NakanoNishimura2026} analyze exact nonlinear Domar aggregation in inelastic production networks 
and uncover a fundamental asymmetry: while positive sectoral shocks exhibit mutual offsetting effects, adverse sectoral shocks reinforce one another through \textit{negative synergy}, creating the possibility that a sufficiently severe negative shock triggers a \textit{negative singularity} in which general equilibrium prices diverge.

In this paper, we bridge exact nonlinear network aggregation, structural empirical calibration, and global dynamic general equilibrium modeling to quantify the cumulative macroeconomic and welfare cost of input-output rigidity. Our analysis makes three main contributions.

First, on the theoretical front, we formalize the mechanism of \textit{network impedance} and \textit{mean sinking} while explicitly accounting for the boundary of negative equilibrium singularities. Because an inelastic CES network admits a strictly positive equilibrium price vector only when productivity realizations remain above a critical singularity threshold ($X > b$, with $b < 0$), the expected Domar Aggregation Function (DAF) must be evaluated as a conditional expectation over the nonsingularity domain. We prove that although conditioning on $\{X > b\}$ truncates the lower tail and shifts the conditional mean of the microeconomic shock strictly into positive territory ($\mathbb{E}[X \mid X > b] > 0$), this truncation uplift is exponentially small ($o(\tau^2)$) as microeconomic volatility $\tau$ becomes small. Consequently, the second-order concavity penalty governed by the network's input-output topology---which we isolate as the \textit{network impedance} coefficient $\mathscr{C}$---strictly dominates, converting symmetric zero-mean Gaussian microeconomic volatility into a persistent negative drift in aggregate productivity (\textit{mean sinking}). Furthermore, parameterizing the degree to which sectors rely directly on primary value added rather than roundabout intermediate inputs---a structural property we term \textit{network primarity} ($\kappa$)---we prove analytically that deepening intermediate supply chains strictly increases network impedance ($\mathscr{C}'(1) < 0$) and amplifies mean sinking.

Second, on the dynamic quantitative front, we show that static mean sinking is magnified dramatically once embedded in an intertemporal general equilibrium growth environment. In a standard Real Business Cycle (RBC) economy, a persistent negative drift in aggregate productivity innovations compounds through autoregressive persistence $(1 - \rho)^{-1}$ and depresses the expected marginal product of capital, inducing households to accumulate a smaller capital stock. Because standard local perturbation or log-linearization methods impose certainty equivalence and erase both the nonlinear mean-sinking drift and endogenous downside skewness, we solve the household's dynamic programming problem globally via Value Function Iteration (VFI) using the exact general equilibrium DAF distribution. This dynamic amplification implies that network rigidity imposes a permanent structural level loss that conventional countercyclical stabilization policies---which aim merely to dampen variance around a given trend---cannot offset.

Third, we bring this framework directly to the data using the U.S.\ Bureau of Economic Analysis (BEA) $174$-sector Input-Output Accounts ($t_0 = 2012$). Employing a non-parametric dual T\"{o}rnqvist calibration strategy, we estimate unrestricted sector-specific elasticities of substitution $\hat{\sigma}_i$ alongside sector-specific AR(1) productivity innovation volatilities $\hat{\tau}_i$ and aggregate price-dual TFP persistence ($\hat{\rho} = 0.5315$). Our empirical estimates confirm pervasive input complementarity across U.S.\ industries: the cross-sectoral mean elasticity is $0.4939$, with $86.2\%$ of sectors exhibiting $\hat{\sigma}_i < 1$ and $14.4\%$ exhibiting non-positive point estimates ($\hat{\sigma}_i \leq 0$). To ensure that retaining unrestricted negative elasticities does not compromise numerical tractability, we prove that transforming the general equilibrium pricing system into logarithmic price coordinates yields a strict contraction mapping in the $\ell_\infty$ norm for arbitrary real elasticities $\sigma_i \in \mathbb{R}$, guaranteeing uniqueness and global convergence.

Feeding the calibrated $174$-sector volatility vector $\widehat{\boldsymbol{\tau}}$ into our exact general equilibrium contraction mapping generates striking quantitative results. Even though underlying microeconomic log-productivity innovations are strictly Gaussian and zero-mean, nonlinear propagation through the empirical U.S.\ production network produces an annual aggregate productivity drift of $-2.32\%$ ($\mathbb{E}[\varepsilon_t] = -0.0232$) alongside pronounced negative skewness ($-0.3455$). Within our annual global RBC model, this mean-sinking penalty lowers stochastic steady-state aggregate TFP by $4.83\%$ and contracts steady-state capital and output by $7.13\%$. Evaluated in terms of Consumption Equivalent Variation (CEV), empirical network impedance inflicts a permanent welfare loss of $-7.10\%$ of aggregate consumption relative to the frictionless Cobb-Douglas benchmark---nearly two orders of magnitude larger than the classical business-cycle welfare cost calculated by \citet{Lucas1987}. Finally, counterfactual experiments varying network primarity ($\kappa \in \{0.9, 1.0, 1.1\}$) demonstrate a sharp structural dichotomy: under Cobb-Douglas technology, deepening supply chains widens welfare dispersion without altering expected welfare (confirming certainty equivalence), whereas under empirical input complementarity, a $10\%$ compression in primary-factor shares deepens the permanent consumption loss monotonically from $-7.10\%$ to $-7.42\%$.

The remainder of the paper is organized as follows. Section~2 develops the static multisectoral general equilibrium framework, derives the analytical properties of network impedance and primarity in a one-sector specialization, and formulates the global dynamic RBC optimization problem. Section~3 describes the BEA $174$-sector dataset and presents our empirical estimates of sectoral elasticities, microeconomic volatilities, and macroeconomic persistence. Section~4 reports the baseline quantitative simulation and welfare results, and Section~5 examines sensitivity to network primarity across Cobb-Douglas and empirical CES regimes. Section~6 concludes with implications for supply-chain resilience and macroeconomic policy. Mathematical proofs of conditional moment bounds, multisectoral impedance scaling, and global $\ell_\infty$ contractivity are collected in Appendices~A--C.
\section{The Model}
\subsection{Multisectoral General Equilibrium}
We begin by formulating a static multisectoral general equilibrium framework to characterize how idiosyncratic sectoral productivity shocks propagate through input-output linkages and aggregate into macroeconomic fluctuations. The economy consists of a representative household and $n$ competitive production sectors, alongside a primary factor of production (e.g., labor or value added).

The representative household maximizes a constant relative risk aversion (CRRA) utility function with risk-aversion parameter $\gamma > 1$, defined over a Cobb-Douglas aggregate of $n$ sectoral consumption goods $(c_1, \dots, c_n)$. To nest logarithmic utility smoothly as $\gamma \to 1$ and maintain consistency with the dynamic formulation in Section~2.3, the household's preferences in direct and indirect utility forms read as follows:
\begin{align}
{V} = 
\frac{\left( \prod_{j=1}^n (c_j/\mu_j)^{\mu_j} \right)^{1-\gamma} - 1}{1-\gamma}
= \frac{\left( {M} / \prod_{j=1}^n \left( {p_j} \right)^{\mu_j} \right)^{1-\gamma} - 1}{1-\gamma},
\end{align}
where ${M}$ denotes the aggregate nominal income of the household, $p_j$ denotes the price of the $j$th good, and $\mu_j > 0$ represents the consumption expenditure weight satisfying $\sum_{j=1}^n \mu_j = 1$. Note that $C \equiv \prod_{j=1}^n (c_j/\mu_j)^{\mu_j}$ defines the real aggregate consumption (or final output) bundle, while $P \equiv \prod_{j=1}^n (p_j)^{\mu_j}$ represents the corresponding ideal price index. Since nominal expenditure satisfies ${M} = CP$, normalizing the primary factor endowment such that ${M} = 1$ implies that aggregate real output growth coincides exactly with aggregate total factor productivity (TFP) growth. In logarithmic terms, this aggregate index is measured by the negative log change in the aggregate price index as follows:
\begin{align}
\log C = - \log P = - \sum_{j=1}^n \mu_j \log p_j = - \boldsymbol{\mu} \log \mathbf{p},
\end{align}
where $\boldsymbol{\mu} = (\mu_1, \dots, \mu_n)$ is the row vector of final expenditure shares and $\log \mathbf{p} = (\log p_1, \dots, \log p_n)^{\top}$ is the column vector of log sectoral prices. Moreover, applying Roy's identity to the indirect utility function confirms that the household's expenditure share on each good $j$ is constant and equal to $\mu_j$:
\begin{align}
\frac{p_j c_j}{{M}} = - \frac{p_j}{{M}}\frac{\partial {V}/ \partial p_j}{\partial {V} / \partial {M}} = \mu_j
&&
(j = 1, \dots, n).
\end{align}

On the production side, each sector $i \in \{1, \dots, n\}$ operates under constant returns to scale and perfect competition, combining the primary factor (indexed by $0$) and intermediate inputs from all $n$ sectors. Sectoral production technology is dualized by a constant elasticity of substitution (CES) unit cost function $g_i$ and a Hicks-neutral productivity level $z_i > 0$. Under zero-profit equilibrium conditions, the output price $p_i$ equals the minimum unit cost:
\begin{align}
{p}_i 
&= ({z}_i)^{-1} {g}_i(p_0, p_1, \dots, p_n \mid \sigma_i)  \notag
\\ 
&=({z}_i)^{-1} \left( 
\alpha_{i0} (p_0)^{1-\sigma_i}
+ \alpha_{i1} (p_1)^{1-\sigma_i}
+ \cdots + \alpha_{in} (p_n)^{1-\sigma_i} 
\right)^{\frac{1}{1-\sigma_i}},
\label{production_i}
\end{align}
where $\sigma_i > 0$ denotes the sector-specific elasticity of substitution across inputs, and the distribution parameters satisfy $\alpha_{ij} \geq 0$ with $\sum_{j=0}^{n} \alpha_{ij} = 1$ for every $i = 1, \dots, n$. For later benchmark comparisons, we also specify the Cobb-Douglas limiting case obtained as $\sigma_i \to 1$, under which the unit cost function takes the log-linear form:
\begin{align}
g_i(p_0, p_1, \dots, p_n \mid 1)
= (p_0)^{\alpha_{i0}} (p_1)^{\alpha_{i1}} \cdots (p_n)^{\alpha_{in}}.
\end{align}

Choosing the primary factor as the numeraire ($p_0 = 1$), the system of sectoral pricing equations \eqref{production_i} can be written compactly in vector notation as a fixed-point problem:
\begin{align}
\mathbf{p} = \langle \mathbf{z}^{-1} \rangle \mathbf{g}(\mathbf{p} \mid \boldsymbol{\sigma})
\qquad
\Longrightarrow
\qquad
\mathbf{p} = \mathbf{h}(\mathbf{z} \mid \boldsymbol{\sigma}),
\label{GEP}
\end{align}
where $\boldsymbol{\sigma} = (\sigma_1, \dots, \sigma_n)$ denotes the vector of sectoral elasticities, $\mathbf{g} = (g_1, \dots, g_n)^{\top}$, and angle brackets $\langle \cdot \rangle$ denote diagonalization of a vector into a diagonal matrix. Note that the operator $\langle \mathbf{z}^{-1} \rangle \mathbf{g}: \mathbf{p} \to \mathbf{p}$ is a contraction mapping that globally converges to a unique fixed point whenever one exists within its domain \citep{Krasnoselskii1964}, \textit{provided} that all unit cost functions $(g_1, \dots, g_n)$ are concave in $\mathbf{p}$.\footnote{However, since the positive orthant $\mathbf{p} \in \mathbb{R}_{+}^{n}$ is unbounded, $\langle \mathbf{z}^{-1} \rangle \mathbf{g}$ does not map a compact set into itself; consequently, the existence of a strictly positive fixed point is not unconditionally guaranteed for arbitrary productivity realizations.} Concavity of $g_i$ is strictly satisfied whenever $\sigma_i > 0$. Whenever a positive fixed point exists, Equation \eqref{GEP} defines an implicit equilibrium pricing mapping $\mathbf{h}$ that translates the vector of microeconomic productivity levels $\mathbf{z}$ into the general equilibrium price vector $\mathbf{p}$. Substituting this equilibrium mapping into the aggregate output identity yields our central analytical object, the \textit{Domar Aggregation Function} (DAF):
\begin{align}
- \boldsymbol{\mu} \log \mathbf{p} 
= - \boldsymbol{\mu} \log \mathbf{h} (\mathbf{z} \mid \boldsymbol{\sigma})
\equiv \mathscr{D}(\log \mathbf{z} \mid \boldsymbol{\sigma}).
\end{align}

When all sectors share a universal elasticity of substitution, $\sigma_i = \sigma$ for all $i$, the fixed-point system \eqref{GEP} admits a closed-form matrix solution. Letting $\mathbf{v} = (\alpha_{10}, \dots, \alpha_{n0})^{\top}$ denote the column vector of primary factor input shares and $\mathbf{A} = [\alpha_{ij}]_{i,j=1}^n$ denote the $n \times n$ matrix of intermediate input coefficients, Equation \eqref{GEP} becomes
\begin{align}
\mathbf{p}
=
\langle \mathbf{z}^{-1} \rangle
\left(
\mathbf{v} + \mathbf{A} \mathbf{p}^{1-\sigma}
\right)^{\frac{1}{1-\sigma}},
\end{align}
where $\mathbf{p}^{1-\sigma} = (p_1^{1-\sigma}, \dots, p_n^{1-\sigma})^{\top}$ denotes element-wise exponentiation. Raising both sides to the power $1-\sigma$ and solving the resulting linear system for $\mathbf{p}^{1-\sigma}$ gives $\mathbf{p}^{1-\sigma} = [\langle e^{(1-\sigma)\log \mathbf{z}} \rangle - \mathbf{A}]^{-1} \mathbf{v}$. Consequently, the universal-elasticity DAF takes the explicit closed form:
\begin{align}
\mathscr{D}(\log \mathbf{z} \mid \sigma)
= - \frac{1}{1-\sigma} 
\boldsymbol{\mu} \log \left( \left[ \langle e^{(1-\sigma)\log \mathbf{z}} \rangle
- \mathbf{A} \right]^{-1} \mathbf{v} \right).
\end{align}
Finally, taking the limit $\sigma \to 1$ recovers the linear Cobb-Douglas benchmark:
\begin{align}
\mathscr{D}(\log \mathbf{z} \mid 1)
= \boldsymbol{\mu} \left[ \mathbf{I} - \mathbf{A} \right]^{-1}
\log \mathbf{z}.
\end{align}
In this frictionless Cobb-Douglas limiting case, Hulten's theorem holds globally, and the row vector $\boldsymbol{\mu}\left[ \mathbf{I} - \mathbf{A} \right]^{-1}$ coincides with the classical Domar weights (the ratios of sectoral gross output to aggregate GDP). 
\subsection{Stochastic Domar Aggregation }
To isolate the fundamental mechanism through which input-output complementarity distorts macroeconomic aggregation under uncertainty, we first analyze a transparent one-sector specialization of our framework; the full multisectoral generalization is developed in Appendix~\ref{A_B}.

Consider a one-sector CES production economy ($n = 1$) in which output is produced using a primary factor with share $v \in (0, 1)$ and its own output as an intermediate input with share $a \in (0, 1)$, such that $v + a = 1$. Under the normalization $p_0 = 1$, the equilibrium price equation reduces to
\begin{align}
p = z^{-1} \left( v + {a} p^{1-\sigma} \right)^{\frac{1}{1-\sigma}}.
\end{align}
Solving this scalar fixed-point equation for $p$ and substituting into $-\mu \log p$ yields the closed-form one-sector Domar Aggregation Function:
\begin{align}
\mathscr{D}(\log z \mid \sigma) = - \frac{1}{1-\sigma} \mu \log \left(  \left[ e^{(1-\sigma) \log z} - {a} \right]^{-1} v \right).
\end{align}
We now evaluate the expected aggregate productivity under stochastic microeconomic fluctuations, assuming that log productivity follows a zero-mean Gaussian distribution, $\log z \sim \mathcal{N}(0, \tau^2)$, where $\tau > 0$ measures microeconomic volatility. Taking expectations of the DAF gives
\begin{align}
\mathbb{E} \left[ \mathscr{D}(\log z \mid \sigma) \right]
= \frac{1}{1-\sigma}\mu \mathbb{E}\left[ \log \left( e^{(1-\sigma) \log z} - {a} \right) - \log v 
\right].
\label{EDAF}
\end{align}

For convenience, let $X \equiv \log z \sim \mathcal{N}(0, \tau^2)$. Notice that the logarithmic term inside \eqref{EDAF} is real-valued and finite if and only if $e^{(1-\sigma)X} > a$, or equivalently $X > \frac{\log {a}}{1-\sigma} \equiv {b}$. When $\sigma < 1$, we have $b < 0$; if a severe negative productivity shock violates $X > b$, the intermediate input requirement exceeds the economy's productive capacity and the equilibrium price diverges to infinity (a \textit{negative singularity}). Thus, the unconditional expectation is indeterminate unless we condition on the \textit{nonsingularity condition} $\{X > b\}$. We therefore evaluate the conditional expectation:
\begin{align}
\mathbb{E}\left[ \log \left( e^{(1-\sigma) X} - {a} \right)
\mid X > {b} \right],
&&
X \sim \mathcal{N}(0, \tau^2).
\end{align}
Expanding $\log(e^{(1-\sigma)X} - a)$ in a 
Maclaurin series 
and taking expectations conditional on $\{X > b\}$ yields
\begin{align}
\mathbb{E}\left[ \log \left( e^{(1-\sigma)X} - {a} \right)
\mid X > {b} \right]
&=  
\log (1 - {a})
+ \frac{1-\sigma}{1-{a}} \mathbb{E}[X \mid X > {b}]
- \frac{1}{2} \frac{(1-\sigma)^2 {a}}{(1-{a})^2} \mathbb{E}[X^2 \mid X > {b}]
+ \cdots \notag
\\ 
&= 
\log v
- \frac{1}{2} {a}{(1-{a})^{-2}}(1-\sigma)^2 \tau^2 + o(\tau^2),
\end{align}
where the second equality invokes $1 - a = v$ together with the asymptotic conditional moment approximations $\mathbb{E}[X \mid X > b] = o(\tau^2)$, $\mathbb{E}[X^2 \mid X > b] = \tau^2 + o(\tau^2)$, and $\mathbb{E}[X^k \mid X > b] = o(\tau^2)$ for all $k \geq 3$, as established in Appendix~\ref{A_A}. Substituting this expansion back into \eqref{EDAF}, the $\log v$ terms cancel out and we obtain the second-order expected DAF:
\begin{align}
\mathbb{E} \left[ \mathscr{D}(\log z \mid \sigma) \mid 
z^{1-\sigma} > {a}
\right]
= - \frac{1}{2} \mu 
\underbrace{
{a}{(1-{a})^{-2}}
}_{=\mathscr{C}}
(1-\sigma) \tau^2 + o(\tau^2).
\end{align}
This expression reveals a striking macroeconomic penalty: whenever inputs are gross complements ($\sigma < 1$), symmetric zero-mean microeconomic shocks ($\mathbb{E}[\log z] = 0$) induce a strictly negative shift in expected aggregate productivity—a phenomenon we term \textit{mean sinking}. The structural multiplier $\mathscr{C} \equiv a(1-a)^{-2} > 0$, which we refer to as \textit{network impedance} coefficient, governs the intensity with which microeconomic variance $\tau^2$ is converted into an aggregate productivity loss.

To examine how the depth of intermediate input linkages shapes network impedance, we introduce a scaling parameter $\kappa \in (0, (1-a)^{-1})$ and define the transformed primary and intermediate input shares as $\tilde{v} = \kappa v$ and $\tilde{a} = 1 - \kappa v$, respectively. We refer to $\kappa$ as the \textit{primarity} of the production network. Setting $\kappa = 1$ recovers the reference state ($\tilde{v} = v$, $\tilde{a} = a$), whereas lowering $\kappa$ reduces reliance on primary factors and deepens roundabout intermediate production; in the limit $\kappa \to 0$, we have $\tilde{a} \to 1$, representing an ultimate circular economy. Expressing network impedance as a function of primarity $\kappa$ gives
\begin{align}
\mathscr{C}(\kappa) = \frac{\tilde{a}}{(1 - \tilde{a})^2} = \frac{1}{\kappa^2 (1 - {a})^2} - \frac{1}{\kappa (1 - {a})} > 0.
\end{align}
Differentiating $\mathscr{C}(\kappa)$ with respect to $\kappa$ over the admissible domain $0 < \kappa < (1-a)^{-1}$ shows that
\begin{align}
\mathscr{C}^\prime(\kappa) = \frac{\kappa(1-{a}) - 2}{\kappa^3 (1-{a})^2}
< \frac{1-2}{\kappa^3 (1-{a})^2} < 0.
\end{align}
Thus, network impedance is strictly decreasing in primarity $\kappa$: deepening the supply chain (reducing $\kappa$) amplifies the macroeconomic mean-sinking penalty nonlinearly.

Finally, it is instructive to contrast these results with the Cobb-Douglas benchmark ($\sigma = 1$) under parameterized primarity $\kappa$. In this log-linear case, the DAF simplifies to
\begin{align}
\mathscr{D}(\log z \mid 1, \kappa)
= - \mu (1 - \tilde{a})^{-1} \log z
= - \mu ((1 - a) \kappa)^{-1} \log z,
\end{align}
whose first two unconditional moments are given by
\begin{align}
\mathbb{E}[\mathscr{D}(\log z \mid 1, \kappa)] = 0,
&&
\Var \left( \mathscr{D}(\log z \mid 1, \kappa) \right)
= \left( \frac{\mu}{(1 - a) \kappa} \right)^2 \tau^2.
\end{align}
In a Cobb-Douglas economy, the expected aggregate shock remains identically zero regardless of the network's primarity $\kappa$; changing $\kappa$ alters only the aggregate volatility through the Domar multiplier $((1-a)\kappa)^{-1}$. By contrast, under empirically relevant input complementarity ($\sigma < 1$), lowering 
$\kappa$ not only magnifies aggregate volatility but also depresses the mean level of aggregate productivity itself. 
\subsection{Macroeconomic Dynamics and Welfare}

Having characterized how idiosyncratic sectoral productivity shocks propagate through the input-output network and aggregate via the Domar Aggregation Function (DAF) into a macroeconomic productivity innovation $\varepsilon_t \equiv \mathscr{D}(\log \mathbf{z}_t \mid \boldsymbol{\sigma}, \kappa)$, we now embed this endogenous shock process into a dynamic stochastic general equilibrium growth framework. This allows us to evaluate how static network impedance shapes intertemporal capital accumulation, precautionary saving, and lifetime social welfare.

Time is discrete and indexed by $t = 0, 1, 2, \dots$. Given initial aggregate capital $K_0 > 0$ and initial aggregate total factor productivity $A_0 > 0$, the representative household chooses a state-contingent sequence of aggregate consumption $\{C_t\}_{t=0}^{\infty}$ and next-period capital $\{K_{t+1}\}_{t=0}^{\infty}$ to maximize expected discounted lifetime utility:
\begin{maxi!}
    {C_0, C_1, \cdots}          
    {\mathcal{V}= \mathbb{E} \sum_{t=0}^{\infty} \beta^t U(C_t) \label{obj}} 
    {\label{prb}}  
    {} %
    \addConstraint K_{t+1} = F(A_t, K_t) - C_t \label{const}
    \addConstraint \log A_t = \rho \log A_{t-1} + \varepsilon_t, \qquad \varepsilon_t 
    \sim \mathscr{D}(\log \mathbf{z}_t \sim \mathcal{N}(0, \boldsymbol{\tau}^2)
  \mid  \boldsymbol{\sigma}, \kappa) \label{law_A}
\end{maxi!}
where $\beta \in (0, 1)$ denotes the subjective discount factor and $\mathbb{E}$ denotes the mathematical expectation operator conditional on information at $t = 0$. In the shock-generating process \eqref{law_A}, the underlying microeconomic log-productivity innovations $\log \mathbf{z}_t = (\log z_{1t}, \dots, \log z_{nt})^\intercal$ are assumed to follow independent zero-mean Gaussian distributions with sector-specific volatilities $\boldsymbol{\tau} = (\tau_1, \dots, \tau_n)^\intercal$; that is, $\log z_{it} \sim \mathcal{N}(0, \tau_i^2)$ independently across sectors $i = 1, \dots, n$ and over time.\footnote{
(Microfoundation note) Specifically, interpreting the primary factor composite in Section~2.1 as a Cobb-Douglas bundle $M_t = K_t^\theta L^{1-\theta}$ with inelastic labor $L = 1$ and normalizing its composite price index to unity in every period ($p_{0t} = 1$) implies that real aggregate output equals $Y_t = (p_{0t} M_t)/P_t = A_t K_t^\theta$, where constant returns to scale ensure that relative prices $\mathbf{p}_t$—and hence $\log A_t = -\boldsymbol{\mu}\log \mathbf{p}_t$—are invariant to the capital stock $K_t$.
}
 Consistent with the static preferences in Section~2.1, the period utility function $U(C_t)$ takes the normalized CRRA form:
\begin{align}
U(C_t) = \frac{1}{1 - \gamma} \left( C_t^{1-\gamma} - 1 \right), 
&&
\gamma > 1,
\end{align}
which preserves the relative risk aversion parameter $\gamma > 1$ while smoothly nesting logarithmic utility $U(C_t) = \log C_t$ in the limit $\gamma \to 1$.

In the 
constraint \eqref{const}, $F(A_t, K_t)$ represents total available output net of capital depreciation:
\begin{align}
F(A_t, K_t) = A_t K_t^\theta + (1 - \delta) K_t,
\end{align}
where $\theta \in (0, 1)$ is the capital output elasticity (with inelastic primary labor normalized to unity) and $\delta \in (0, 1)$ is the physical depreciation rate of capital. In the law of motion \eqref{law_A}, aggregate log productivity $\log A_t$ follows a stationary first-order autoregressive process with persistence parameter $\rho \in (0, 1)$ (distinct from the Perron-Frobenius spectral radius in Appendix~\ref{A_B}). Crucially, rather than imposing an exogenous symmetric Gaussian innovation as in standard Real Business Cycle (RBC) models, the macroeconomic shock $\varepsilon_t$ is drawn independently across periods from the exact general-equilibrium DAF distribution $\mathscr{D}(\log \mathbf{z}_t \sim \mathcal{N}(0, \boldsymbol{\tau}^2) \mid \boldsymbol{\sigma}, \kappa)$ induced by heterogeneous sectoral volatilities $\boldsymbol{\tau}$, sectoral elasticities $\boldsymbol{\sigma}$, and network primarity $\kappa$ (conditional on nonsingularity). Under empirical input complementarity ($\sigma_i < 1$), $\varepsilon_t$ exhibits both a negative structural drift (\textit{mean sinking}, $\mathbb{E}[\varepsilon_t] < 0$) and negative skewness.

Letting primes denote next-period variables, the dynamic optimization problem \eqref{prb} admits the recursive Bellman representation over the continuous state space $(K, A) \in \mathbb{R}_+^2$:
\begin{align}
\mathcal{V}(K, A) = \max_{0 < C < F(A, K)} \left\{ U(C) + \beta \, \mathbb{E}_{\varepsilon'} \left[ \mathcal{V}\!\left(F(A, K) - C, \, \exp(\rho \log A + \varepsilon')\right) \right] \right\},
\label{bellman}
\end{align}
where $\mathbb{E}_{\varepsilon'}[\cdot]$ denotes the expectation taken over the empirical DAF innovation distribution $\mathscr{D}(\log \mathbf{z}' \sim \mathcal{N}(0, \boldsymbol{\tau}^2) \mid \boldsymbol{\sigma}, \kappa)$. Standard local perturbation or log-linearization methods around a certainty-equivalent steady state rely on first-order certainty equivalence and therefore eliminate both the nonlinear mean-sinking bias and higher-order downside tail risks induced by network rigidity. To capture the full nonlinear interplay between input-output impedance, precautionary capital accumulation, and macroeconomic curvature, we solve the Bellman equation \eqref{bellman} globally via Value Function Iteration (VFI) over a discretized two-dimensional grid, evaluating the conditional expectation via Monte Carlo integration over simulated DAF realizations. Let $C^*(K, A)$ and $K'^*(K, A) = F(A, K) - C^*(K, A)$ denote the resulting optimal consumption and capital policy functions.

Finally, to quantify the aggregate welfare cost of network impedance in transparent economic units, we compute the \textit{Consumption Equivalent Variation} ($\lambda_{\mathrm{CEV}}$). Let $\mathcal{V}^{\mathrm{CD}}$ and $\mathcal{V}^{\mathrm{CES}}(\boldsymbol{\sigma}, \kappa)$ denote the unconditional expected discounted lifetime utility generated along 
stochastic simulations under the frictionless Cobb-Douglas benchmark ($\boldsymbol{\sigma} = \mathbf{1}, \kappa = 1$) and the rigid CES production network, respectively. We define $\lambda_{\mathrm{CEV}}$ as the permanent proportional adjustment in consumption across all dates and states required to leave a household in the Cobb-Douglas benchmark indifferent to living in the rigid CES economy:
\begin{align}
\mathbb{E}\sum_{t=0}^{\infty} \beta^t U\!\left((1 + \lambda_{\mathrm{CEV}}) C_t^{\mathrm{CD}}\right) = \mathcal{V}^{\mathrm{CES}}(\boldsymbol{\sigma}, \kappa).
\end{align}
Exploiting the homogeneity of the normalized CRRA utility function yields the 
closed-form welfare metric:
\begin{align}
\lambda_{\mathrm{CEV}} = \left( \frac{\mathcal{V}^{\mathrm{CES}}(\boldsymbol{\sigma}, \kappa)(1 - \beta)(1 - \gamma) + 1}{\mathcal{V}^{\mathrm{CD}}(1 - \beta)(1 - \gamma) + 1} \right)^{\frac{1}{1 - \gamma}} - 1.
\label{cev_formula}
\end{align}
A negative value ($\lambda_{\mathrm{CEV}} < 0$) measures the permanent percentage loss in aggregate consumption equivalent to the welfare burden imposed by network rigidity, jointly encompassing the steady-state productivity depression from mean sinking and the dynamic risk premium from asymmetric tail shocks. 
\section{Empirical Modeling}
To bring our multisectoral general equilibrium framework to the data, we calibrate the structural parameters $(\boldsymbol{\sigma}, \boldsymbol{\tau}, \rho)$ using historical input-output accounts for the United States. In this section, we detail the dataset construction, our dual calibration methodology, and the resulting empirical distribution of the sector-specific elasticities of substitution $\boldsymbol{\sigma} = (\sigma_1, \dots, \sigma_n)^\intercal$.

\subsection{Sectoral Elasticities}

\subsubsection{Data Source and Numeraire Normalization}

Our empirical analysis draws on the annual Benchmark and Summary Input-Output Accounts published by the U.S.\ Bureau of Economic Analysis (BEA), disaggregated into $n = 174$ private and public production sectors. Specifically, we utilize both nominal and real (chain-type quantity index) Use tables alongside nominal and real Gross Output tables across all available sample years $t \in \{1, \dots, T\}$, designating $t_0 = 2012$ as the reference base year.

For each sector $i \in \{1, \dots, n\}$ and year $t$, let $Z_{jit}^{\mathrm{nom}}$ denote the nominal intermediate expenditure by sector $i$ on inputs produced by sector $j \in \{1, \dots, n\}$, let $\textit{VA}_{it}^{\mathrm{nom}}$ and $\textit{VA}_{it}^{\mathrm{real}}$ denote nominal and real value added in sector $i$, and let $\textit{GO}_{it}^{\mathrm{nom}}$ and $\textit{GO}_{it}^{\mathrm{real}}$ denote nominal and real gross output, respectively. The time-varying nominal cost shares of intermediate inputs and primary value added ($j = 0$) in sector $i$ are constructed as
\begin{align}
s_{ijt} = \frac{Z_{jit}^{\mathrm{nom}}}{\textit{GO}_{it}^{\mathrm{nom}}} \quad (j = 1, \dots, n),
&&
\text{and}
&&
s_{i0t} = \frac{\textit{VA}_{it}^{\mathrm{nom}}}{\textit{GO}_{it}^{\mathrm{nom}}},
\end{align}
which satisfy the accounting identity $\sum_{j=0}^{n} s_{ijt} = 1$ for every sector $i$ and year $t$. Setting $t = t_0$ recovers the base-year cost-share parameters $\alpha_{ij} \equiv s_{ij, t_0}$ entering the CES unit cost function \eqref{production_i}, as well as the baseline intermediate input matrix $\mathbf{A} = [\alpha_{ij}]_{i,j=1}^n$ and primary-factor vector $\mathbf{v} = (\alpha_{10}, \dots, \alpha_{n0})^\intercal$.

To align observed nominal prices with the theoretical numeraire convention ($p_{0t} = 1$) established in Section~2.1, we first compute raw sectoral gross output deflators $\tilde{p}_{it} = \textit{GO}_{it}^{\mathrm{nom}} / \textit{GO}_{it}^{\mathrm{real}}$ and the aggregate economy-wide value-added deflator $\tilde{p}_{0t} = \left(\sum_{i=1}^n \textit{VA}_{it}^{\mathrm{nom}}\right) / \left(\sum_{i=1}^n \textit{VA}_{it}^{\mathrm{real}}\right)$. Dividing each sectoral price index by $\tilde{p}_{0t}$ and normalizing all prices to unity in the base year $t_0$ yields the relative price series:
\begin{align}
p_{it} = \frac{\tilde{p}_{it} / \tilde{p}_{0t}}{\tilde{p}_{i,t_0} / \tilde{p}_{0,t_0}} \quad (i = 1, \dots, n),
&&
\text{and}
&&
p_{0t} = 1 
\end{align}
This dual normalization ensures that all equilibrium price movements reflect genuine relative cost shifts with respect to the primary factor numeraire.

\subsubsection{Dual T\"{o}rnqvist Calibration 
}

Calibrating sector-level elasticities of substitution across $174$ interconnected inputs poses well-known dimensional and conceptual challenges. Rather than relying on parametric instrumental-variable (IV) regressions, we adopt a non-parametric \textit{dual T\"{o}rnqvist calibration} strategy that matches observed unit-cost movements directly to a superlative price index.

In index number theory, the Sato--Vartia index (Sato, 1976; Vartia, 1976) is known to be exact for a two-input (or single-elasticity) CES aggregator. However, once production involves multiple heterogeneous inputs ($n = 174$) generated by an arbitrary smooth homothetic cost function, the Sato--Vartia index is no longer exact for a multi-input CES specification and reduces to a second-order approximation. Because exactness is unattainable in general multi-input production networks regardless of whether Sato-Vartia or T\"{o}rnqvist weights are employed, we align our CES unit cost function with the \textit{T\"{o}rnqvist index}---the canonical superlative index (Diewert, 1976) widely established in growth accounting and productivity measurement.

Under a general twice-differentiable (translog-equivalent) unit cost structure, the dual T\"{o}rnqvist index provides a superlative measure of sector $i$'s Hicks-neutral total factor productivity (TFP) level $z_{it}^{\mathrm{TL}}$ relative to the base year $t_0 = 2012$ ($z_{i,t_0}^{\mathrm{TL}} = 1$). In logarithmic form, sectoral TFP is recovered directly from observed relative output and input prices together with average nominal cost shares:
\begin{align}
\log z_{it}^{\mathrm{TL}} = -\log p_{it} + \sum_{j=0}^{n} \bar{s}_{ijt} \log p_{jt},
&&
\text{where}
&&
\bar{s}_{ijt} \equiv \frac{1}{2}\left(s_{ijt} + \alpha_{ij}\right).
\label{eq:tornqvist_tfp}
\end{align}
By the zero-profit pricing condition \eqref{production_i}, observed output price and dual productivity jointly imply an empirical \textit{target unit cost} ${g}_{it}^{\mathrm{target}} \equiv p_{it} z_{it}^{\mathrm{TL}} = \exp\bigl(\sum_{j=0}^n \bar{s}_{ijt} \log p_{jt}\bigr)$ that the parametric CES cost specification must replicate in year $t$.

Accordingly, for each sector $i \in \{1, \dots, n\}$ and each non-base year $t \neq t_0$, we identify the period-specific effective elasticity of substitution $\hat{\sigma}_{it}$ by matching the theoretical CES unit cost function ${g}_i(\mathbf{p}_t \mid \sigma)$—evaluated at the base-year cost shares $\alpha_{ij}$ and observed relative prices $\mathbf{p}_t = (1, p_{1t}, \dots, p_{nt})^\intercal$—to the target unit cost ${g}_{it}^{\mathrm{target}}$:
\begin{align}
\hat{\sigma}_{it} = \arg\min_{\sigma \in \mathbb{R}} 
\left(
\left( \sum\nolimits_{j 
} 
\alpha
_{ij} \, p_{jt}^{1-\sigma} 
\right)^{\frac{1}{1-\sigma}} - {g}_{it}^{\mathrm{target}} 
\right)^2,
\label{eq:sigma_opt}
\end{align}
where the problem has a unique zero solution.\footnote{In the numerical implementation, Equation~\eqref{eq:sigma_opt} is solved via the quasi-Newton L-BFGS-B algorithm with a function tolerance of $10^{-15}$, smoothly switching to the Cobb-Douglas geometric mean $\prod_{j 
} p_{jt}^{{\alpha}_{ij}}$ whenever $|\sigma - 1| < 10^{-6}$.
} Taking the temporal arithmetic mean across all $T - 1$ non-base sample years yields our time-invariant structural elasticity estimate for sector $i$:
\begin{align}
\hat{\sigma}_i = \frac{1}{T - 1} \sum\nolimits_{t \neq t_0} \hat{\sigma}_{it}.
\label{eq:sigma_mean}
\end{align}

Two methodological considerations justify matching observed equilibrium price-share co-movements directly without instrumental-variable filtering. First, whereas causal inference aimed at isolating partial-equilibrium demand or supply slopes requires instruments to disentangle simultaneous shifts, our objective is \textit{general equilibrium simulation}: we seek to faithfully reproduce the realized joint behavior of sectoral prices and cost shares in the actual economy. Filtering out equilibrium price co-variation would distort the empirical transmission mechanism that our model is designed to replicate. Second, from a theoretical standpoint, our constant-returns-to-scale CES production structure rests on the maintained assumption of perfect competition, under which individual producers act as price takers. Positing reverse causality from a sector's input demand to input prices in order to motivate endogeneity corrections would therefore be internally inconsistent with the competitive price-taking foundation of the model.

\subsubsection{Empirical Distribution and Implications for Network Equilibrium}

Table~\ref{tab:sigma_summary} summarizes the cross-sectional distribution of the estimated sectoral elasticities $\hat{\sigma}_i$ across all $174$ BEA industries. Two salient empirical findings emerge.

\begin{table}[htbp]
\centering
\caption{Summary Statistics of Estimated Sectoral Substitution Elasticities ($\hat{\sigma}_i$, $n = 174$)}
\label{tab:sigma_summary}
\begin{threeparttable}
\begin{tabular}{l S[table-format=1.4] S[table-format=1.4] S[table-format=1.4] S[table-format=1.4] S[table-format=1.4]}
\toprule
{Parameter} & {Mean} & {Std.\ Dev.} & {Median} & {25th Pct.} & {75th Pct.} \\
\midrule
$\hat{\sigma}_i$ (All Sectors, $n = 174$) & 0.4939 & 0.5494 & 0.4245 & 0.2190 & 0.7131 \\
\bottomrule
\end{tabular}
\begin{tablenotes}[flushleft]
\footnotesize
\item \textit{Notes}: Estimates are computed from the U.S.\ BEA 174-sector Input-Output Accounts using base year $t_0 = 2012$ and value added as numeraire ($p_{0t} = 1$). Across all $174$ industries, $86.2\%$ exhibit gross input complementarity ($\hat{\sigma}_i < 1$), and $14.4\%$ exhibit non-positive point estimates ($\hat{\sigma}_i \leq 0$). Each sector's structural elasticity $\hat{\sigma}_i$ represents the temporal mean of annual dual T\"{o}rnqvist-matched CES elasticities $\hat{\sigma}_{it}$ defined in Equations~\eqref{eq:sigma_opt}--\eqref{eq:sigma_mean}.
\end{tablenotes}
\end{threeparttable}
\end{table}

First, the cross-sectional distribution is heavily concentrated below the Cobb-Douglas benchmark of unity: the mean elasticity is $0.4939$ (with a median of $0.4245$ and an interquartile range of $[0.2190, 0.7131]$), and $86.2\%$ of all U.S.\ industries exhibit gross input complementarity ($\hat{\sigma}_i < 1$). In light of our analytical results in Section~2.2 and Appendix~\ref{A_B}, this pervasive complementarity confirms that the empirical U.S.\ production network possesses substantial network impedance, converting symmetric microeconomic volatility into aggregate mean sinking.

Second, because the calibration problem \eqref{eq:sigma_opt} is solved over the unrestricted real line ($\sigma \in \mathbb{R}$), $14.4\%$ of industries exhibit negative point estimates ($\hat{\sigma}_i \leq 0$). Although a negative substitution elasticity renders the sectoral unit cost function $g_i(\mathbf{p} \mid \hat{\sigma}_i)$ convex rather than concave in price levels $\mathbf{p} \in \mathbb{R}_+^n$—seemingly departing from standard neoclassical cost-minimization axioms—consistent with our simulation philosophy, if such price-share responses are empirically observed in the historical accounts, they should be preserved and reproduced in the quantitative model rather than artificially censored. Crucially, Appendix~\ref{A_C} proves that the equilibrium pricing mapping remains a strict contraction in logarithmic price coordinates ($\mathbf{x} = \log \mathbf{p} \in \mathbb{R}^n$) with respect to the $\ell_\infty$ norm for arbitrary real elasticities $\sigma_i \in \mathbb{R}$. We therefore retain the full unrestricted vector of empirical estimates $\widehat{\boldsymbol{\sigma}} = (\hat{\sigma}_1, \dots, \hat{\sigma}_{174})^\intercal$ in our general equilibrium simulations without compromising uniqueness or global numerical convergence.
\subsection{Microeconomic Volatilities}

To parameterize the stochastic shock-generating process in Equation~\eqref{law_A}, we next estimate the sector-specific volatility vector $\boldsymbol{\tau} = (\tau_1, \dots, \tau_n)^\intercal$ governing the dispersion of idiosyncratic productivity innovations across all $n = 174$ BEA industries.

\subsubsection{Time-Series Extraction of Sectoral Innovation Volatilities}

Using the annual series of sectoral relative prices $\{p_{it}\}_{t=1}^T$ and nominal cost shares $\{s_{ijt}\}_{t=1}^T$ constructed in Section~3.1, we first compute the historical panel of dual T\"{o}rnqvist total factor productivity levels $z_{it}^{\mathrm{TL}}$ via Equation~\eqref{eq:tornqvist_tfp} for each industry $i \in \{1, \dots, n\}$ and sample year $t \in \{1, \dots, T\}$. Taking natural logarithms yields the longitudinal sectoral log-productivity series $\log z_{it}^{\mathrm{TL}}$.

Because our dynamic general equilibrium framework explicitly models macroeconomic persistence through the aggregate state transition equation \eqref{law_A}, the microeconomic inputs fed into the Domar Aggregation Function must represent serially uncorrelated period-by-period innovations rather than cumulative level changes. Accordingly, for each sector $i \in \{1, \dots, n\}$, our baseline specification estimates a first-order autoregressive process on log sectoral TFP by ordinary least squares:
\begin{align}
\log z_{it}^{\mathrm{TL}} = c_i + \varrho_i \log z_{i,t-1}^{\mathrm{TL}} + u_{it},
\label{eq:sector_ar1}
\end{align}
where $c_i$ is a sector-specific intercept, $\varrho_i$ captures sectoral persistence, and $u_{it}$ denotes the idiosyncratic productivity innovation. We then measure sector $i$'s baseline microeconomic volatility $\hat{\tau}_i^{\mathrm{AR1}}$ by the sample standard deviation of the estimated AR(1) residuals $\{\hat{u}_{it}\}$:
\begin{align}
\hat{\tau}_i^{\mathrm{AR1}} = \sqrt{\frac{1}{T - 2} \sum_{t=2}^{T} \left(\hat{u}_{it} - \bar
{\hat{u}}_i\right)^2},
\end{align}
where $\bar
{\hat{u}}_i$ denotes the sample mean of the residuals.

For comparison and robustness, we also compute an alternative volatility measure under the assumption that sectoral log productivity follows a geometric Brownian motion, taking the sample standard deviation of annual log-differences $\Delta \log z_{it}^{\mathrm{TL}} \equiv \log z_{it}^{\mathrm{TL}} - \log z_{i,t-1}^{\mathrm{TL}}$:
\begin{align}
\hat{\tau}_i^{\mathrm{GBM}} = \sqrt{\frac{1}{T - 2} \sum_{t=2}^{T} \left(\Delta \log z_{it}^{\mathrm{TL}} - \overline{
{\Delta \log z}_i^{\mathrm{TL}}}\right)^2}.
\end{align}
Removing the autoregressive component in \eqref{eq:sector_ar1} prevents double-counting persistence across microeconomic and macroeconomic layers while isolating pure annual productivity surprises.

\subsubsection{Cross-Sectional Heterogeneity in Sectoral Volatility}

Table~\ref{tab:tau_summary} reports the cross-sectional summary statistics for both volatility estimators across all $174$ U.S.\ industries. Under our baseline AR(1) innovation specification, the cross-sectoral mean volatility is $\bar{\tau}^{\mathrm{AR1}} = 0.0254$ ($2.54\%$ per annum) with a median of $0.0182$ and an interquartile range spanning from $0.0137$ to $0.0278$. The log-difference estimator $\hat{\tau}_i^{\mathrm{GBM}}$ exhibits a closely aligned distribution, with a mean of $0.0273$ and a median of $0.0191$, confirming that persistent level drift accounts for only a modest fraction of annual sectoral TFP variation.

\begin{table}[htbp]
\centering
\caption{Summary Statistics of Estimated Sectoral Productivity Volatilities ($\hat{\tau}_i$, $n = 174$)}
\label{tab:tau_summary}
\begin{threeparttable}
\begin{tabular}{l S[table-format=1.4] S[table-format=1.4] S[table-format=1.4] S[table-format=1.4] S[table-format=1.4] 
}
\toprule
{Estimator} & {Mean} & {Std.\ Dev.} & {Median} & {25th Pct.} & {75th Pct.} 
\\
\midrule
$\hat{\tau}_i^{\mathrm{AR1}}$ (AR(1) Innovation) & 0.0254 & 0.0242 & 0.0182 & 0.0137 & 0.0278 
\\
$\hat{\tau}_i^{\mathrm{GBM}}$ (Log-Difference)   & 0.0273 & 0.0267 & 0.0191 & 0.0145 & 0.0283 
\\
\bottomrule
\end{tabular}
\begin{tablenotes}[flushleft]
\footnotesize
\item \textit{Notes}: Volatilities are estimated from annual dual T\"{o}rnqvist sectoral log-TFP panels ($\log z_{it}^{\mathrm{TL}}$) across $n = 174$ U.S.\ BEA industries. $\hat{\tau}_i^{\mathrm{AR1}}$ denotes the sample standard deviation of residuals from sector-level AR(1) regressions, whereas $\hat{\tau}_i^{\mathrm{GBM}}$ denotes the sample standard deviation of first log-differences.
\end{tablenotes}
\end{threeparttable}
\end{table}

A notable feature of Table~\ref{tab:tau_summary} is the pronounced right-skewness and dispersion across sectors: the cross-sectional standard deviation ($0.0242$) is nearly as large as the mean, and sectoral innovation volatilities range over more than an order of magnitude from a minimum of $0.0054$ ($0.54\%$) in stable service industries to a maximum of $0.2344$ ($23.44\%$) in primary commodity and energy-related sectors. As demonstrated in Appendix~\ref{A_B}, because mean productivity sinking scales quadratically with sectoral volatility ($\tau_i^2$) weighted by network impedance, accurately capturing this cross-sectional heterogeneity via the full empirical vector $\widehat{\boldsymbol{\tau}} = (\hat{\tau}_1^{\mathrm{AR1}}, \dots, \hat{\tau}_{174}^{\mathrm{AR1}})^\intercal$ is essential for quantifying aggregate welfare losses.
\subsection{Macroeconomic Persistence}

The final empirical parameter required to close the dynamic general equilibrium model is the autoregressive persistence parameter $\rho \in (0, 1)$ governing the propagation of aggregate total factor productivity in Equation~\eqref{law_A}.

\subsubsection{Aggregate Productivity Reconstruction and AR(1) Estimation}

Consistent with our theoretical aggregation identity in Section~2.1---where aggregate real output and aggregate productivity under normalized primary endowments ($M = 1$) coincide with the negative logarithm of the ideal consumption price index---we construct the historical annual series of aggregate log productivity $\log A_t^{\mathrm{emp}}$ directly from observed sectoral relative prices $\{p_{jt}\}_{j=1}^n$ and base-year final demand expenditure shares $\{\mu_j\}_{j=1}^n$:
\begin{align}
\log A_t^{\mathrm{emp}} = -\log P_t = -\sum_{j=1}^{n} \mu_j \log p_{jt} = -\boldsymbol{\mu} \log \mathbf{p}_t,
\label{eq:agg_tfp_emp}
\end{align}
covering the full sample period through $2025$. Constructing $\log A_t^{\mathrm{emp}}$ via \eqref{eq:agg_tfp_emp} ensures complete methodological consistency between the historical series used to estimate dynamic persistence and the general equilibrium Domar Aggregation Function $\mathscr{D}(\log \mathbf{z}_t \mid \boldsymbol{\sigma}, \kappa) = -\boldsymbol{\mu} \log \mathbf{p}_t$ simulated in our counterfactual experiments.

To separate business-cycle fluctuations from low-frequency secular productivity trends, we apply the standard Hodrick--Prescott (HP) filter to $\log A_t^{\mathrm{emp}}$ and extract the cyclical component $\tilde{a}_t \equiv \log A_t^{\mathrm{emp}} - \tau_t^{\mathrm{HP}}$, where $\tau_t^{\mathrm{HP}}$ denotes the smooth trend component. We then estimate a first-order autoregressive process on the detrended cyclical series by ordinary least squares:
\begin{align}
\tilde{a}_t = \rho \, \tilde{a}_{t-1} + \eta_t,
\label{eq:macro_ar1}
\end{align}
where $\eta_t$ represents the reduced-form cyclical productivity innovation.

Estimating Equation~\eqref{eq:macro_ar1} on the U.S.\ BEA series yields an autoregressive persistence parameter of
\begin{align}
\hat{\rho} = 0.5315,
\end{align}
with a reduced-form innovation standard deviation of $\hat{\sigma}_\eta = 0.0054$. Notably, an annual persistence of $\hat{\rho} = 0.5315$ corresponds to a quarterly autocorrelation of approximately $0.5315^{1/4} \approx 0.8538$ (or roughly $0.78$ after accounting for temporal aggregation of time-averaged annual flows), placing our direct annual estimate squarely within the standard range of the Real Business Cycle literature while maintaining exact frequency alignment with the annual BEA Input-Output Accounts. Moreover, although the relatively short annual sample limits statistical power for higher-moment tests, the estimated AR(1) residuals $\{\hat{\eta}_t\}$ exhibit a pronounced negative skewness of $-0.57$, closely aligning in both sign and magnitude with the endogenous downside skewness ($-0.35$) generated by our $174$-sector CES general equilibrium simulations in Section~4.

\subsubsection{Summary of Empirical Calibration}

Taken together, Sections~3.1--3.3 provide a fully data-disciplined parameterization $(\widehat{\boldsymbol{\sigma}}, \widehat{\boldsymbol{\tau}}, \hat{\rho})$ derived from a unified accounting framework. In Section~4, rather than imposing an exogenous symmetric Gaussian innovation with variance $\hat{\sigma}_\eta^2$ as in standard representative-agent models, we feed the heterogeneous microeconomic volatility vector $\widehat{\boldsymbol{\tau}}$ and unrestricted sectoral elasticity vector $\widehat{\boldsymbol{\sigma}}$ into the exact general equilibrium pricing system \eqref{GEP}. This generates the endogenous, negatively skewed macroeconomic shock distribution $\mathscr{D}(\log \mathbf{z}_t \sim \mathcal{N}(0, \widehat{\boldsymbol{\tau}}^2) \mid \widehat{\boldsymbol{\sigma}}, \kappa)$, which we then combine with the annual persistence $\hat{\rho} = 0.5315$ to solve the global dynamic programming problem and evaluate aggregate social welfare.
\section{RBC and Welfare}
Having calibrated the structural elasticity vector $\widehat{\boldsymbol{\sigma}}$, the sector-specific innovation volatility vector $\widehat{\boldsymbol{\tau}}$, and the macroeconomic persistence parameter $\hat{\rho}$ from the U.S.\ BEA Input-Output Accounts, we now quantify the dynamic macroeconomic and welfare consequences of input-output network impedance. Specifically, we solve the global dynamic programming problem \eqref{bellman} and compare the resulting ergodic consumption and welfare distributions between the frictionless Cobb-Douglas benchmark ($\boldsymbol{\sigma} = \mathbf{1}, \kappa = 1.0$) and the empirically calibrated CES production network ($\widehat{\boldsymbol{\sigma}}, \kappa = 1.0$).

\subsection{Quantitative Calibration and Global Numerical Solution}

To maintain complete frequency alignment with the annual BEA Input-Output Accounts and our direct annual persistence estimate $\hat{\rho} = 0.5315$, we calibrate all structural parameters of the dynamic growth model at an annual frequency. Table~\ref{tab:calibration_params} summarizes the baseline parameter values. The subjective discount factor is set to $\beta = 0.96$, corresponding to an annual risk-free real interest rate of approximately $4.17\%$. The annual physical capital depreciation rate is set to $\delta = 0.10$, the capital output elasticity is calibrated to $\theta = 0.33$, and the household's relative risk aversion coefficient is set to $\gamma = 2.0$, all of which conform to standard quantitative macroeconomic conventions.

\begin{table}[htbp]
\centering
\caption{Baseline Annual Calibration Parameters for the Dynamic RBC Model}
\label{tab:calibration_params}
\begin{threeparttable}
\begin{tabular}{llcl}
\toprule
{ } & {Description} & {Value} & {Source / Target} \\
\midrule
$\beta$     & Subjective discount factor       & $0.9600$ & Standard annual calibration ($\approx 4\%$ real rate) \\
$\delta$    & Physical depreciation rate       & $0.1000$ & Standard annual capital depreciation ($10\%$ p.a.) \\
$\gamma$    & Relative risk aversion           & $2.0000$ & Standard CRRA curvature \\
$\theta$    & Capital output elasticity        & $0.3300$ & Standard aggregate capital income share \\
$\hat{\rho}$ & Aggregate TFP persistence       & $0.5315$ & Estimated from BEA price-dual TFP (Section~3.3) \\
$\widehat{\boldsymbol{\sigma}}$ & Sectoral CES elasticities & {Vector} & Dual T\"{o}rnqvist calibration (Section~3.1) \\
$\widehat{\boldsymbol{\tau}}$   & Sectoral shock volatilities & {Vector} & Sectoral AR(1) residuals (Section~3.2) \\
\bottomrule
\end{tabular}
\begin{tablenotes}[flushleft]
\footnotesize
\item \textit{Notes}: All dynamic parameters are calibrated at an annual frequency to match the annual BEA 174-sector Input-Output Accounts ($t_0 = 2012$).
\end{tablenotes}
\end{threeparttable}
\end{table}

Our quantitative solution proceeds in three stages. First, we draw $10,000$ independent realizations of the $174$-dimensional microeconomic log-productivity shock vector $\log \mathbf{z}_t \sim \mathcal{N}(\mathbf{0}, 
\widehat{\boldsymbol{\tau}}^2 
)$ and solve the exact general equilibrium pricing system \eqref{GEP} via contraction mapping (guaranteed to converge globally by Appendix~\ref{A_C}) for each realization, thereby constructing the empirical Domar Aggregation Function (DAF) distribution $\mathscr{D}(\log \mathbf{z}_t \sim \mathcal{N}(\mathbf{0}, \widehat{\boldsymbol{\tau}}^2) \mid \widehat{\boldsymbol{\sigma}}, \kappa)$. Second, we solve the Bellman equation \eqref{bellman} globally via Value Function Iteration (VFI) over a discretized two-dimensional state grid of $N_K = 100$ capital points spanning $[0.5 K_{ss}, 1.5 K_{ss}]$ and $N_A = 50$ log-spaced productivity points spanning $[\log A_{ss} - 3\sigma_{\log A}, \log A_{ss} + 3\sigma_{\log A}]$, using bivariate spline interpolation and $1,000$ Monte Carlo shock draws to evaluate conditional expectations up to a uniform value convergence tolerance of $10^{-6}$. Third, using the resulting global policy functions $C^*(K, A)$ and $K'^*(K, A)$, we simulate $1,000$ ergodic stochastic economies of length $T = 500$ years (discarding an initial burn-in of $100$ years) via exact bootstrapping with common random numbers across scenarios to evaluate lifetime utility $\mathcal{V}$ and the Consumption Equivalent Variation $\lambda_{\mathrm{CEV}}$ defined in Equation~\eqref{cev_formula}.

\subsection{Aggregate Shock Asymmetry and Mean Sinking}

Panel~A of Table~\ref{tab:welfare_baseline} contrasts the summary moments of the general equilibrium macroeconomic productivity innovation $\varepsilon_t$ across the $10,000$ Monte Carlo realizations under the Cobb-Douglas benchmark ($\boldsymbol{\sigma} = \mathbf{1}$) and the empirical CES network ($\widehat{\boldsymbol{\sigma}}$).

\begin{table}[htbp]
\centering
\caption{Macroeconomic Shock Moments, Steady States, and Welfare Cost ($\kappa = 1.0$)}
\label{tab:welfare_baseline}
\begin{threeparttable}
\begin{tabular}{l S[table-format=-2.4] S[table-format=-2.4]}
\toprule
{Metric} & {Cobb-Douglas ($\boldsymbol{\sigma} = \mathbf{1}$)} & {Empirical CES ($\widehat{\boldsymbol{\sigma}}$)} \\
\midrule
\multicolumn{3}{l}{\textit{Panel A: General Equilibrium DAF Shock Moments ($\varepsilon_t$, $10,000$ draws)}} \\
Mean ($\mathbb{E}[\varepsilon_t]$)          & -0.0001 & -0.0232 \\
Variance ($\Var(\varepsilon_t)$)          &  0.0001 &  0.0001 \\
Standard Deviation ($\mathrm{SD}(\varepsilon_t)$) & 0.0109 & 0.0111 \\
Skewness                                    & -0.0047 & -0.3455 \\
\midrule
\multicolumn{3}{l}{\textit{Panel B: Stochastic Steady-State Levels}} \\
Aggregate TFP ($A_{ss}$)                    &  0.9990 &  0.9507 \\
Capital Stock ($K_{ss}$)                    &  3.5277 &  3.2763 \\
\midrule
\multicolumn{3}{l}{\textit{Panel C: Ergodic Social Welfare ($1,000$ paths, $T = 500$)}} \\
Mean Lifetime Utility ($\mathbb{E}[\mathcal{V}]$) & 3.5052 & 1.8632 \\
Std.\ Dev.\ of Lifetime Utility ($\mathrm{SD}(\mathcal{V})$) & 0.1029 & 0.1145 \\
Consumption Equivalent Variation ($\lambda_{\mathrm{CEV}}$, \%) & {--} & -7.10 \\
\bottomrule
\end{tabular}
\begin{tablenotes}[flushleft]
\footnotesize
\item \textit{Notes}: Both economies share identical microeconomic innovation distributions $\log \mathbf{z}_t \sim \mathcal{N}(\mathbf{0}, \widehat{\boldsymbol{\tau}}^2)$ and baseline network primarity $\kappa = 1.0$. Exact unrounded DAF shock means and variances are $-0.000060$ and $0.000118$ for Cobb-Douglas, and $-0.023201$ and $0.000123$ for Empirical CES. The exact CEV loss is $-7.097\%$.
\end{tablenotes}
\end{threeparttable}
\end{table}

Although the underlying sectoral productivity innovations $\log z_{it}$ are strictly symmetric and centered at zero, nonlinear propagation through the empirical U.S.\ production network fundamentally alters the distribution of aggregate shocks. Under the frictionless Cobb-Douglas specification, Hulten's theorem holds globally and the sample mean of $\varepsilon_t$ is indistinguishable from zero ($\mathbb{E}[\varepsilon_t] = -0.000060$), with near-zero skewness ($-0.0047$) and an annual standard deviation of $1.09\%$. By contrast, once empirical input complementarity ($\hat{\sigma}_i < 1$ for $86.2\%$ of sectors) is introduced, the expected macroeconomic innovation drops sharply to $\mathbb{E}[\varepsilon_t] = -0.023201$—an annual \textit{mean-sinking} penalty of $-2.32\%$—while the distribution develops pronounced negative skewness ($-0.3455$). Intuitively, because intermediate inputs are gross complements, adverse productivity shocks in bottleneck sectors cannot be easily substituted away and generate steep nonlinear price spikes, whereas favorable sectoral shocks are dampened by rigid complementary input requirements.

\subsection{Steady-State Compression and Dynamic Welfare Loss}

\begin{figure}[htbp]
\centering
\includegraphics[width=0.82\textwidth]{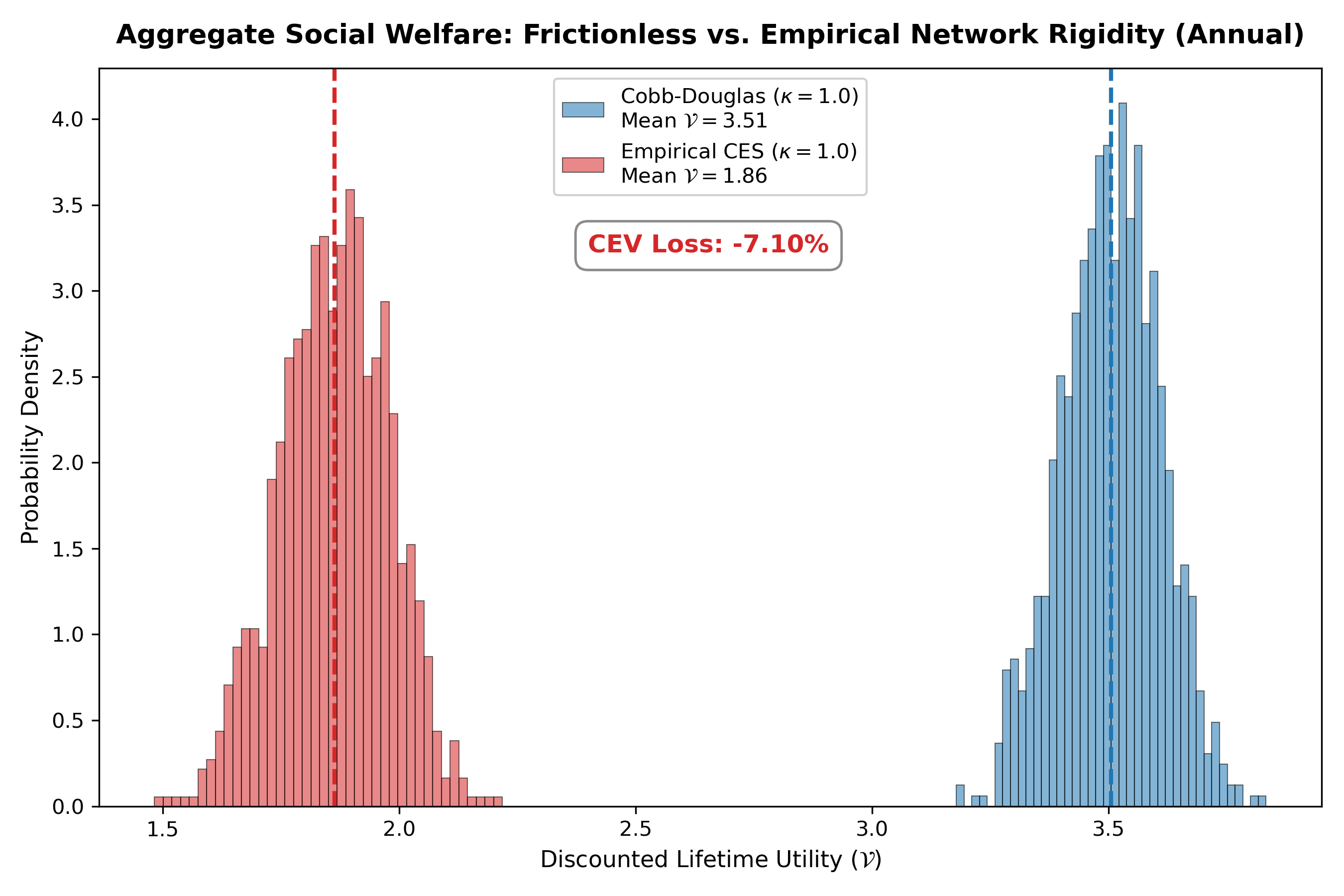}
\caption{Ergodic Distributions of Social Welfare: Cobb-Douglas Benchmark vs.\ Empirical CES Network ($\kappa = 1.0$)}
\label{fig:welfare_cd_emp}
\end{figure}

Panels~B and~C of Table~\ref{tab:welfare_baseline} report the resulting macroeconomic steady states and dynamic welfare outcomes, while Figure~\ref{fig:welfare_cd_emp} displays the full ergodic probability distributions of discounted lifetime utility $\mathcal{V}$ across the $1,000$ simulated economies.

The quantitative impact of network impedance is substantial and operates through two reinforcing dynamic channels. First, the $-2.32\%$ annual mean-sinking drift compounds through macroeconomic persistence $(1 - \hat{\rho})^{-1} = (1 - 0.5315)^{-1} \approx 2.13$, depressing the stochastic steady-state aggregate TFP level $A_{ss} = \exp\bigl(\mathbb{E}[\varepsilon_t]/(1 - \hat{\rho})\bigr)$ by $4.83\%$, from $0.9990$ in the Cobb-Douglas economy to $0.9507$ in the empirical CES economy. Second, within the neoclassical growth mechanism, a lower expected return on capital induces households to scale back capital accumulation: steady-state capital $K_{ss}$ contracts by $7.13\%$ (from $3.5277$ to $3.2763$), amplifying the initial TFP depression by the capital multiplier $(1 - \theta)^{-1} \approx 1.49$.

As illustrated in Figure~\ref{fig:welfare_cd_emp}, the empirical CES welfare distribution shifts dramatically to the left with zero overlap against the Cobb-Douglas benchmark: expected discounted lifetime utility falls from $\mathcal{V}^{\mathrm{CD}} = 3.5052$ to $\mathcal{V}^{\mathrm{CES}} = 1.8632$, while welfare dispersion increases by $11.3\%$ (from a standard deviation of $0.1029$ to $0.1145$). Converted into permanent consumption units via Equation~\eqref{cev_formula}, moving from a frictionless Cobb-Douglas production structure to the empirical U.S.\ input-output network inflicts a welfare loss equivalent to a **permanent $-7.10\%$ ($-7.097\%$) reduction in aggregate consumption**.

In his seminal benchmark calculation, Lucas (1987) demonstrated that eliminating aggregate consumption fluctuations in a representative-agent economy yields a welfare gain of less than $0.1\%$ of consumption, because symmetric fluctuations around a fixed trend carry only a second-order risk cost. Subsequent static multisectoral studies (e.g., Baqaee and Farhi, 2019) showed that supply-chain nonlinearities generate first-order output losses by degrading average productivity. Our dynamic general equilibrium results reveal that when empirical multisectoral CES nonlinearities interact with endogenous intertemporal capital accumulation and downside tail risk, the resulting welfare cost of network rigidity is nearly two orders of magnitude larger than Lucas's classical business-cycle cost.
\section{Sensitivity Analysis}
Our analytical derivations in Section~2.2 and Appendix~\ref{A_B} establish a sharp theoretical dichotomy regarding how supply-chain depth governs macroeconomic outcomes. Under a frictionless Cobb-Douglas network ($\sigma_i = 1$), reducing network \textit{primarity} $\kappa$ magnifies the variance of aggregate productivity shocks through larger Domar multipliers while leaving expected aggregate productivity invariant at zero. Under an empirically inelastic CES network ($\sigma_i < 1$), by contrast, lowering $\kappa$ strictly increases network impedance ($\mathscr{C}'(1) < 0$), simultaneously amplifying aggregate volatility and deepening the structural \textit{mean-sinking} penalty. In this section, we test both theoretical predictions quantitatively on the full $174$-sector U.S.\ input-output network.

\subsection{Parameterizing Supply-Chain Depth across Network Regimes}

Following Equation~\eqref{eq:A_kappa} in Appendix~\ref{A_B}, we construct counterfactual input-output architectures by scaling the baseline primary value-added share vector $\mathbf{v} = (\alpha_{10}, \dots, \alpha_{n0})^\intercal$ by the primarity parameter $\kappa > 0$, setting $\tilde{\mathbf{v}}(\kappa) = \kappa \mathbf{v}$ and adjusting intermediate input shares proportionally via $\mathbf{A}(\kappa) = \langle \mathbf{1} - \kappa \mathbf{v} \rangle \mathbf{P}$, where $\mathbf{P} = \langle \mathbf{1} - \mathbf{v} \rangle^{-1} \mathbf{A}$ is the row-stochastic intermediate allocation matrix. By construction, $\kappa = 1.0$ recovers the observed 2012 U.S.\ input-output structure.

To evaluate comparative statics around the empirical status quo while remaining strictly inside the nonsingularity domain, we examine three primarity regimes under both the Cobb-Douglas benchmark ($\boldsymbol{\sigma} = \mathbf{1}$) and the empirical CES specification ($\widehat{\boldsymbol{\sigma}}$):
\begin{enumerate}
    \item \textbf{Rigid / Deepened Supply Chain ($\kappa = 0.9$)}: Primary value-added shares are compressed by $10\%$ across all sectors ($\tilde{v}_i = 0.9 v_i$), increasing intermediate input intensity and lengthening roundabout supply chains (higher circularity and larger Domar multipliers).
    \item \textbf{Baseline Network ($\kappa = 1.0$)}: Observed U.S.\ BEA input-output shares ($\tilde{v}_i = v_i$).
    \item \textbf{Flexible / Modular Supply Chain ($\kappa = 1.1$)}: Primary value-added shares are expanded by $10\%$ across all sectors ($\tilde{v}_i = 1.1 v_i$), reducing intermediate input roundaboutness and lowering network multipliers.
\end{enumerate}
For each of the six resulting specifications $(\boldsymbol{\sigma}, \kappa) \in \{\mathbf{1}, \widehat{\boldsymbol{\sigma}}\} \times \{0.9, 1.0, 1.1\}$, we hold the microeconomic volatility vector $\widehat{\boldsymbol{\tau}}$ and all dynamic RBC parameters fixed at their baseline values from Table~\ref{tab:calibration_params}, re-solving the $10,000$ general equilibrium DAF realizations, the global Bellman equation \eqref{bellman}, and the $1,000$ ergodic simulation paths.

\subsection{Certainty Equivalence vs.\ Impedance Distortion}

\begin{figure}[htbp]
\centering
\includegraphics[width=\textwidth]{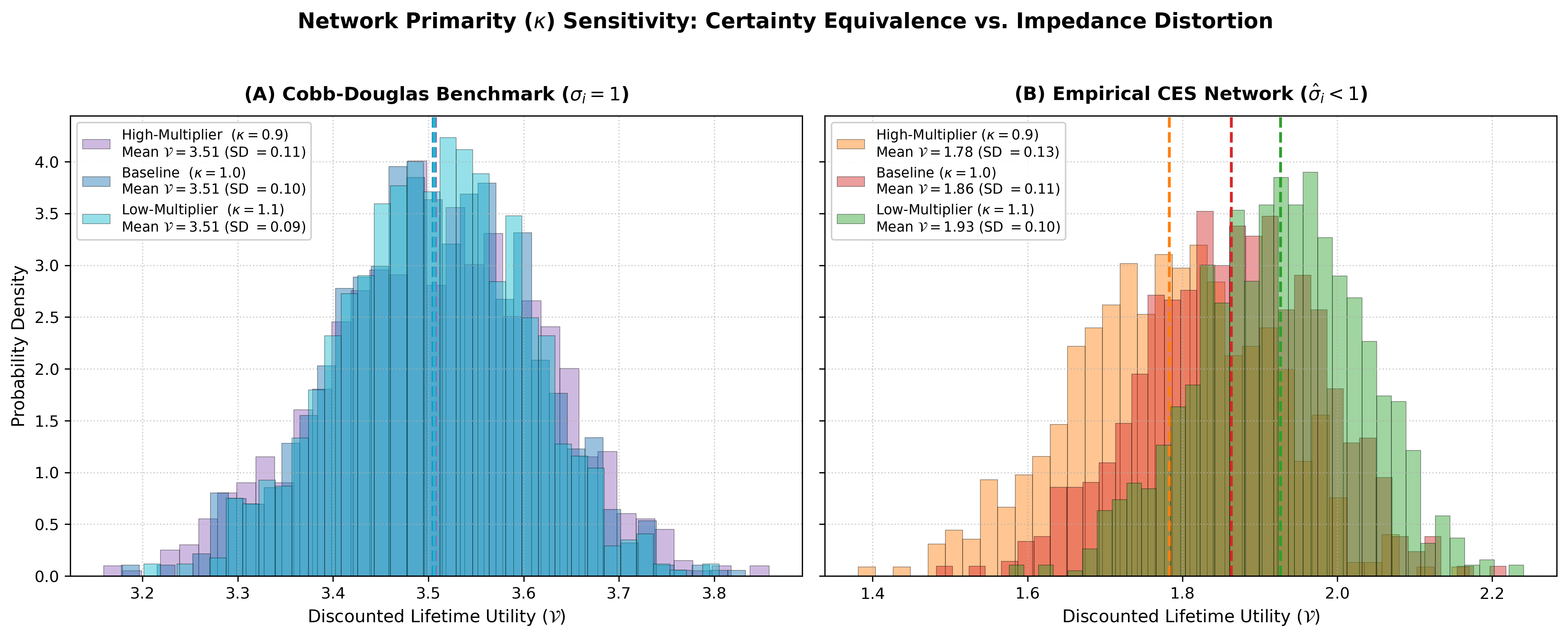}
\caption{Network Primarity ($\kappa$) Sensitivity: Certainty Equivalence under Cobb-Douglas (Panel~A) vs.\ Impedance Distortion under Empirical CES (Panel~B)}
\label{fig:welfare_kappa_sidebyside}
\end{figure}

Table~\ref{tab:welfare_sensitivity} summarizes the general equilibrium shock moments, steady-state levels, and dynamic welfare outcomes across all six economies, while Figure~\ref{fig:welfare_kappa_sidebyside} contrasts the ergodic welfare distributions of the Cobb-Douglas benchmark (Panel~A) and the empirical CES network (Panel~B) side by side.

\begin{table}[htbp]
\centering
\caption{Sensitivity of Aggregate Shock Moments and Social Welfare to Network Primarity ($\kappa$)}
\label{tab:welfare_sensitivity}

\resizebox{\textwidth}{!}{%
\begin{tabular}{l S[table-format=-2.4] S[table-format=-2.4] S[table-format=-2.4] S[table-format=-2.4] S[table-format=-2.4] S[table-format=-2.4]}
\toprule
& \multicolumn{3}{c}{{Cobb-Douglas Benchmark ($\boldsymbol{\sigma} = \mathbf{1}$)}} & \multicolumn{3}{c}{{Empirical CES Network ($\widehat{\boldsymbol{\sigma}}$)}} \\
\cmidrule(lr){2-4} \cmidrule(lr){5-7}
{Metric} & {$\kappa = 0.9$} & {$\kappa = 1.0$} & {$\kappa = 1.1$} & {$\kappa = 0.9$} & {$\kappa = 1.0$} & {$\kappa = 1.1$} \\
\midrule
\multicolumn{7}{l}{\textit{Panel A: General Equilibrium DAF Shock Moments ($\varepsilon_t$, $10,000$ draws)}} \\
Mean ($\mathbb{E}[\varepsilon_t]$)               & -0.0001 & -0.0001 & -0.0001 & -0.0243 & -0.0232 & -0.0223 \\
Variance ($\Var(\varepsilon_t)$)                &  0.0001 &  0.0001 &  0.0001 &  0.0002 &  0.0001 &  0.0001 \\
Standard Deviation ($\mathrm{SD}(\varepsilon_t)$) & 0.0119 & 0.0109 & 0.0099 & 0.0122 & 0.0111 & 0.0102 \\
Skewness                                         & -0.0050 & -0.0047 & -0.0043 & -0.3543 & -0.3455 & -0.3360 \\
\midrule
\multicolumn{7}{l}{\textit{Panel B: Stochastic Steady-State Levels}} \\
Aggregate TFP ($A_{ss}$)                         &  0.9998 &  0.9999 &  0.9999 &  0.9495 &  0.9517 &  0.9534 \\
Capital Stock ($K_{ss}$)                         &  3.5321 &  3.5322 &  3.5323 &  3.2700 &  3.2812 &  3.2902 \\
\midrule
\multicolumn{7}{l}{\textit{Panel C: Ergodic Social Welfare ($1,000$ paths, $T = 500$)}} \\
Mean Lifetime Utility ($\mathbb{E}[\mathcal{V}]$) & 3.5075 & 3.5052 & 3.5059 & 1.7831 & 1.8632 & 1.9265 \\
Std.\ Dev.\ of Lifetime Utility ($\mathrm{SD}(\mathcal{V})$) & 0.1137 & 0.1029 & 0.0949 & 0.1265 & 0.1145 & 0.1048 \\
CEV vs.\ Baseline CD ($\lambda_{\mathrm{CEV}}$, \%) & +0.01 & {0.00} & +0.00 & -7.42 & -7.10 & -6.84 \\
\bottomrule
\end{tabular}%
}

\par\smallskip

\begin{minipage}{\textwidth}
\footnotesize
\noindent
\textit{Notes}: All six economies share identical microeconomic innovation distributions $\log \mathbf{z}_t \sim \mathcal{N}(\mathbf{0}, \widehat{\boldsymbol{\tau}}^2)$. Exact unrounded DAF shock means are $-0.000071$, $-0.000060$, and $-0.000050$ for Cobb-Douglas ($\kappa = 0.9, 1.0, 1.1$), and $-0.024267$, $-0.023201$, and $-0.022340$ for Empirical CES ($\kappa = 0.9, 1.0, 1.1$). Exact shock variances are $0.000142$, $0.000118$, and $0.000099$ for Cobb-Douglas, and $0.000150$, $0.000123$, and $0.000104$ for Empirical CES. CEV values are computed relative to the baseline Cobb-Douglas economy ($\boldsymbol{\sigma} = \mathbf{1}, \kappa = 1.0$) via Equation~\eqref{cev_formula}; exact unrounded CEV changes are $+0.0107\%$ ($\mathrm{CD}, \kappa=0.9$), $+0.0035\%$ ($\mathrm{CD}, \kappa=1.1$), $-7.4172\%$ ($\mathrm{CES}, \kappa=0.9$), $-7.0968\%$ ($\mathrm{CES}, \kappa=1.0$), and $-6.8421\%$ ($\mathrm{CES}, \kappa=1.1$).
\end{minipage}

\end{table}


Comparing Panels~A and~B of Figure~\ref{fig:welfare_kappa_sidebyside} reveals a striking structural contrast between frictionless and rigid production networks:

\begin{enumerate}
    \item \textbf{Variance Scaling and Certainty Equivalence in the Cobb-Douglas Economy (Panel~A)}: Under Cobb-Douglas technology ($\boldsymbol{\sigma} = \mathbf{1}$), lowering primarity $\kappa$ from $1.1$ to $0.9$ inflates the variance of aggregate productivity innovations by $43.4\%$ (from $0.000099$ to $0.000142$, raising the annual shock standard deviation from $0.99\%$ to $1.19\%$), which in turn widens the cross-path standard deviation of discounted lifetime utility by $19.8\%$ (from $0.0949$ to $0.1137$). Crucially, however, the three ergodic welfare distributions remain stacked directly on top of one another around $\mathbb{E}[\mathcal{V}] \approx 3.51$, with consumption-equivalent welfare differences of barely $0.004\%$ to $0.011\%$. Thus, in a log-linear network, Lucas's (1987) certainty-equivalence insight holds intact: changes in supply-chain depth alter macroeconomic volatility without materially affecting average social welfare, rendering local linearization methods accurate.
    \item \textbf{Mean Sinking and Cumulative Welfare Gradient in the Empirical CES Economy (Panel~B)}: Once empirical input complementarity ($\widehat{\boldsymbol{\sigma}}$) is introduced, reducing primarity $\kappa$ from $1.1$ to $1.0$ and $0.9$ not only widens aggregate volatility (from a shock standard deviation of $1.02\%$ to $1.22\%$) and intensifies negative skewness (from $-0.3360$ to $-0.3543$), but also systematically shifts the center of the welfare distribution leftward. Expected lifetime utility declines monotonically from $\mathbb{E}[\mathcal{V}] = 1.9265$ ($\kappa = 1.1$) to $1.8632$ ($\kappa = 1.0$) and $1.7831$ ($\kappa = 0.9$), widening the permanent consumption loss relative to the Cobb-Douglas benchmark from **$-6.84\%$** to **$-7.10\%$** and **$-7.42\%$**.
\end{enumerate}

Furthermore, the welfare gradient under empirical CES exhibits clear convexity with respect to supply-chain depth: a $10\%$ compression in primary factor shares ($\kappa: 1.0 \to 0.9$) adds $0.32$ percentage points to the permanent consumption loss (a $4.5\%$ increase in the welfare burden), whereas an equal $10\%$ expansion in primary factor shares ($\kappa: 1.0 \to 1.1$) alleviates the loss by $0.26$ percentage points. This asymmetry mirrors the convex curvature of the impedance function $\mathscr{C}(\kappa)$ derived in Section~2.2 ($\mathscr{C}''(\kappa) > 0$), confirming that the macroeconomic welfare cost of network rigidity materializes specifically through the interaction between input complementarity ($\sigma_i < 1$) and roundabout supply-chain structure.
\section{Conclusion}
This paper has examined how microeconomic productivity fluctuations propagate through empirical input-output linkages to depress long-run macroeconomic performance and social welfare. By integrating exact nonlinear Domar aggregation with a calibrated $174$-sector general equilibrium model of the U.S.\ economy and a global Real Business Cycle growth framework, we demonstrate that input complementarity ($\sigma_i < 1$) fundamentally alters the macroeconomic consequences of sectoral volatility.

Analytically, we show that even when idiosyncratic productivity innovations are symmetric and centered at zero, nonlinear price aggregation across complementary inputs acts as a concave filter that systematically depresses expected aggregate productivity. Evaluating the Domar Aggregation Function conditionally on the nonsingularity domain reveals that this \textit{mean-sinking} penalty is governed by \textit{network impedance}---a structural multiplier determined by the Leontief inverse topology that increases strictly as production relies more heavily on roundabout intermediate inputs rather than primary value added. Quantitatively, calibrating our framework to the U.S.\ BEA Input-Output Accounts reveals that pervasive sectoral complementarity ($\bar{\sigma} = 0.4939$, with $86.2\%$ of industries satisfying $\hat{\sigma}_i < 1$) transforms zero-mean microeconomic Gaussian shocks into an annual macroeconomic TFP drift of $-2.32\%$ with substantial downside skewness. Compounded by empirical macroeconomic persistence ($\hat{\rho} = 0.5315$) and endogenous capital accumulation, this structural drag lowers steady-state output by $7.13\%$ and generates a permanent consumption-equivalent welfare loss of $-7.10\%$ relative to a frictionless Cobb-Douglas economy.

Our findings carry important implications for both macroeconomic modeling and economic policy. Methodologically, our sensitivity analysis across network primarity shows that local linearization and certainty-equivalence approximations remain accurate only when input-output linkages are log-linear; once realistic sectoral complementarity is present, capturing the interaction between network topology, downside tail risk, and precautionary capital accumulation requires global nonlinear solution methods. From a policy perspective, our results suggest that the economic case for enhancing supply-chain resilience extends well beyond insuring against rare catastrophic disruptions. Policies that alleviate production bottlenecks---such as fostering input substitutability, encouraging modular supply-chain design, and diversifying critical intermediate suppliers---reduce structural network impedance, thereby raising the baseline level of aggregate productivity and yielding substantial permanent welfare gains even during normal business-cycle fluctuations.

Several avenues remain open for future research. Extending our global dynamic framework to incorporate endogenous sectoral capital reallocation, inventory buffers, and international trade linkages would further illuminate how economies adapt dynamically to persistent input-output bottlenecks. Investigating how structural transformation and shifts in intermediate input intensity over time alter aggregate network impedance across countries also represents a promising direction for empirical work.
\appendix
\normalsize
\renewcommand{\thesection}{\Alph{section}}
\makeatletter
\renewcommand{\@seccntformat}[1]{Appendix \csname the#1\endcsname\quad}
\makeatother
\numberwithin{equation}{section}

\section{Conditional Moments Evaluation}\label{A_A}
In this appendix, we evaluate the conditional moments of a zero-mean Gaussian productivity shock truncated from below at a negative nonsingularity threshold. As established in Section~2.2, the general equilibrium price vector remains strictly positive and well-defined only when the realization of the log-productivity shock $X = \log z$ exceeds the critical lower bound $b < 0$. Consequently, the expected Domar Aggregation Function (DAF) must be evaluated conditionally on the nonsingularity domain $\{X > b\}$. Below, we demonstrate that as the shock volatility becomes small ($\tau \to 0^+$), the truncation bias captured by the inverse Mills ratio decays exponentially fast, ensuring that the conditional moments coincide with their untruncated Gaussian counterparts up to order $o(\tau^2)$.

Let $X \sim \mathcal{N}(0, \tau^2)$ with probability density $\tau^{-1}\phi(x/\tau)$, where $\phi(\cdot)$ and $\Phi(\cdot)$ denote the standard normal probability density function and cumulative distribution function, respectively, and let $b < 0$ be a fixed negative threshold. Applying the change of variables $u \equiv x/\tau$ and invoking the standard Gaussian identity $u\phi(u) = -\phi'(u)$, the conditional first moment is evaluated as
\begin{align}
\mathbb{E}[X \mid X > b]
&=\frac{1}{\Phi(-b/\tau)}\int_{b}^{\infty} x \phi\left(\frac{x}{\tau}\right) \frac{dx}{\tau}
=\frac{\tau}{\Phi(-b/\tau)}\int_{b/\tau}^{\infty} u \phi(u)\, du
=\frac{\tau}{\Phi(-b/\tau)}\int_{b/\tau}^{\infty} -\phi'(u)\, du 
\notag
\\
&=\frac{\tau}{\Phi(-b/\tau)} \Bigl[ \phi(u) \Bigr]_{\infty}^{b/\tau}
= \tau \frac{\phi(b/\tau)}{\Phi(-b/\tau)} = o(\tau^2).
\end{align}
Because $b < 0$ is fixed, we have $b/\tau \to -\infty$ and $\Phi(-b/\tau) \to 1$ as $\tau \to 0^+$, while $\phi(b/\tau) = (2\pi)^{-1/2}\exp\bigl(-b^2/(2\tau^2)\bigr)$ vanishes exponentially faster than any polynomial power of $\tau$. Hence, the inverse Mills ratio correction term in the conditional mean is strictly of order $o(\tau^2)$.

Next, we turn to the conditional second moment. Integrating by parts with $u\phi(u) = -\phi'(u)$ yields
\begin{align}
\mathbb{E}[X^2 \mid X > b]
&=\frac{1}{\Phi(-b/\tau)}\int_{b}^{\infty} x^2 \phi\left(\frac{x}{\tau}\right) \frac{dx}{\tau}
=\frac{\tau^2}{\Phi(-b/\tau)}\int_{b/\tau}^{\infty} u^2 \phi(u)\, du
\notag
\\
&=\frac{\tau^2}{\Phi(-b/\tau)}
\left(
\Bigl[ u \phi(u) \Bigr]_{\infty}^{b/\tau} + \int_{b/\tau}^{\infty} \phi(u)\, du
\right)
=\frac{\tau^2}{\Phi(-b/\tau)}
\left(
\frac{b}{\tau} \phi(b/\tau) + \Phi(-b/\tau)
\right)
\notag
\\
&= \tau^2 + b \tau \frac{\phi(b/\tau)}{\Phi(-b/\tau)}
= \tau^2 + o(\tau^2).
\end{align}
By the same asymptotic property of the normal tail as $\tau \to 0^+$, the truncation adjustment $b\tau \phi(b/\tau)/\Phi(-b/\tau)$ is exponentially negligible, confirming that $\mathbb{E}[X^2 \mid X > b] = \tau^2 + o(\tau^2)$.

Finally, to bound the higher-order remainder terms $R_3(X)$ in the second-order Maclaurin expansion, consider any integer order $k \geq 3$. Letting $Z \sim \mathcal{N}(0, 1)$ denote a standard normal random variable, we obtain the uniform upper bound
\begin{align}
\mathbb{E}[X^{k} \mid X > b]
&\leq \mathbb{E}[|X|^{k} \mid X > b]
=\frac{1}{\Phi(-b/\tau)}\int_{b}^{\infty} |x|^{k} \phi\left(\frac{x}{\tau}\right) \frac{dx}{\tau}
=\frac{\tau^{k}}{\Phi(-b/\tau)}\int_{b/\tau}^{\infty} |u|^{k} \phi(u)\, du
\notag
\\
&\leq \frac{\tau^{k}}{\Phi(-b/\tau)}\int_{-\infty}^{\infty} |u|^{k} \phi(u)\, du
\leq \frac{\tau^{k}}{1/2} \mathbb{E}[|Z|^{k}] = O(\tau^{k}) = o(\tau^2),
\end{align}
where the last inequality uses $\Phi(-b/\tau) > 1/2$, which follows immediately from $-b/\tau > 0$. Taken together, Equations~(A.1)--(A.3) establish that conditioning on the nonsingularity region $\{X > b\}$ preserves the second-order asymptotic structure of the stochastic Domar aggregation up to $o(\tau^2)$.
\section{Stochastic Domar Aggregation in Multisectoral Networks} \label{A_B}
This appendix generalizes the one-sector stochastic aggregation analysis of Section~2.2 to an arbitrary $n$-sector input-output network. We derive a second-order asymptotic approximation of the multisectoral Domar Aggregation Function (DAF), establish the strict positivity of the resulting network impedance vector under input complementarity, and characterize how changes in supply-chain depth (network primarity) govern both mean productivity sinking and aggregate volatility.

\subsection*{Second-Order Asymptotic Expansion of Equilibrium Prices}

Consider the universal-elasticity multisectoral CES economy introduced in Section~2.1 ($\sigma_i = \sigma$ for all $i = 1, \dots, n$). Let the idiosyncratic log-productivity shock in sector $i$ be denoted by $\ln z_i = {X}_i$, where ${X}_i \sim \text{i.i.d.}\ \mathcal{N}(0, \tau^2)$. Defining the transformed productivity vector as $\mathbf{z}^{1-\sigma} = (e^{{(1-\sigma)} {X}_1}, \dots, e^{{(1-\sigma)} {X}_n})^{\intercal}$, a second-order Maclaurin series expansion of each component $e^{{(1-\sigma)} {X}_i}$ around $X_i = 0$ yields:
\begin{align}
z_i^{(1-\sigma)} = e^{{(1-\sigma)} {X}_i} = 1 + {(1-\sigma)} {X}_i + \frac{1}{2} {(1-\sigma)}^2 {X}_i^2 + {R}_3(X_i),
\end{align}
where $R_3(X_i)$ denotes the third-order Taylor remainder. Arranging these terms along the main diagonal of an $n \times n$ matrix, we obtain the perturbed diagonal productivity operator:
\begin{align}
\langle \mathbf{z}^{1-\sigma} \rangle = \mathbf{I} + {(1-\sigma)} \langle \boldsymbol{X} \rangle + \frac{1}{2} {(1-\sigma)}^2 \langle \boldsymbol{X}^2 \rangle + {R}_3(\boldsymbol{X}) 
= \mathbf{I} + \boldsymbol{\Delta}, \label{eq:z_gamma}
\end{align}
where $\boldsymbol{X} = ({X}_1, \dots, {X}_n)^{\intercal}$, $\boldsymbol{X}^2 = ({X}_1^2, \dots, {X}_n^2)^{\intercal}$, angle brackets indicate diagonalization of a vector, and $\boldsymbol{\Delta} \equiv \langle \mathbf{z}^{1-\sigma} \rangle - \mathbf{I}$ captures the stochastic perturbation matrix.

At the deterministic reference state ($\mathbf{z} = \mathbf{1}$), constant returns to scale imply that the primary factor input coefficient vector is given by $\mathbf{v} = [\mathbf{I} - \mathbf{A}] \mathbf{1}$, where $\mathbf{A}$ denotes the non-negative $n \times n$ matrix of intermediate input coefficients and $\mathbf{1} = (1, \dots, 1)^{\intercal}$. Given a realization of sectoral productivity shocks $\ln \mathbf{z} = \boldsymbol{X}$ within the nonsingularity domain, the general equilibrium price vector $\mathbf{p} = (p_1, \dots, p_n)^{\intercal}$ satisfies $\mathbf{p}^{1-\sigma} = \left[ \langle \mathbf{z}^{1-\sigma} \rangle - \mathbf{A} \right]^{-1} \mathbf{v}$. Applying a Neumann series expansion around $\mathbf{I} - \mathbf{A}$ yields:
\begin{align}
\mathbf{p}^{1-\sigma} 
&= \left[ \langle \mathbf{z}^{1-\sigma} \rangle - \mathbf{A} \right]^{-1} \mathbf{v} 
= \left[ \mathbf{I} - \mathbf{A} + \boldsymbol{\Delta} \right]^{-1} [\mathbf{I} - \mathbf{A}] \mathbf{1} 
= \left( \mathbf{L} - \mathbf{L}\boldsymbol{\Delta} \mathbf{L} + \mathbf{L}\boldsymbol{\Delta} \mathbf{L} \boldsymbol{\Delta} \mathbf{L} - \cdots \right) [\mathbf{I} - \mathbf{A}] \mathbf{1} \nonumber \\
&= \mathbf{1} - \mathbf{L}\boldsymbol{\Delta}\mathbf{1} + \mathbf{L}\boldsymbol{\Delta} \mathbf{L}\boldsymbol{\Delta}\mathbf{1} - \cdots,
\end{align}
where $\mathbf{L} = [\mathbf{I} - \mathbf{A}]^{-1} = [\ell_{ij}]_{i,j=1}^n$ denotes the Leontief inverse matrix. Substituting the perturbation decomposition \eqref{eq:z_gamma} and retaining terms up to second order in $\boldsymbol{X}$, we evaluate the first- and second-order matrix products as:
\begin{align}
\mathbf{L} \boldsymbol{\Delta} 
&= {(1-\sigma)} \mathbf{L} \langle \boldsymbol{X} \rangle + \frac{1}{2} {(1-\sigma)}^2 \mathbf{L} \langle \boldsymbol{X}^2 \rangle + {R}_3(\boldsymbol{X}), 
\\
\mathbf{L} \boldsymbol{\Delta} \mathbf{L} \boldsymbol{\Delta}
&= {(1-\sigma)}^2 \mathbf{L} \langle \boldsymbol{X} \rangle \mathbf{L} \langle \boldsymbol{X} \rangle + {R}_3(\boldsymbol{X}).
\end{align}
Noting that $\langle \boldsymbol{X} \rangle \mathbf{1} = \boldsymbol{X}$ and $\langle \boldsymbol{X}^2 \rangle \mathbf{1} = \boldsymbol{X}^2$, it follows that the transformed equilibrium price vector expands as:
\begin{align}
\mathbf{p}^{1-\sigma} 
=
 \mathbf{1} - {(1-\sigma)} \mathbf{L} \boldsymbol{X} + {(1-\sigma)}^2 \mathbf{L} \langle \boldsymbol{X} \rangle \mathbf{L} \boldsymbol{X} - \frac{1}{2} {(1-\sigma)}^2 \mathbf{L} \boldsymbol{X}^2
+ {R}_3(\boldsymbol{X}).
\end{align}

To translate these price perturbations into logarithmic changes, we apply the second-order Maclaurin expansion of $\ln p_i^{1-\sigma}$ around $p_i^{1-\sigma} = 1$:
\begin{align}
\log {p}_i^{1-\sigma} = ({p}_i^{1-\sigma} - 1) - \frac{1}{2} ({p}_i^{1-\sigma} - 1)^2 + \cdots.
\end{align}
Expressed in vector form across all $i = 1, \dots, n$, where $(\mathbf{p}^{1-\sigma} - \mathbf{1})^2$ denotes element-wise squaring, this expression reduces to:
\begin{align}
\ln \mathbf{p}^{1-\sigma} &= (\mathbf{p}^{1-\sigma} - \mathbf{1}) - \frac{1}{2} (\mathbf{p}^{1-\sigma} - \mathbf{1})^2 + \cdots \nonumber \\
&=
 - {(1-\sigma)} \mathbf{L} \boldsymbol{X} + {(1-\sigma)}^2 \mathbf{L} \langle \boldsymbol{X} \rangle \mathbf{L} \boldsymbol{X} - \frac{1}{2} {(1-\sigma)}^2 \mathbf{L} \boldsymbol{X}^2 - \frac{1}{2} ({(1-\sigma)} \mathbf{L} \boldsymbol{X})^2
+ {R}_3(\boldsymbol{X}).
  \label{eq:ln_pi_gamma}
\end{align}

\subsection*{Multisectoral Network Impedance and Mean Sinking}

To evaluate the conditional expectation of \eqref{eq:ln_pi_gamma}, we expand the two quadratic network interaction vectors element by element:
\begin{align}
\langle \boldsymbol{X} \rangle \mathbf{L} \boldsymbol{X} &=
\begin{bmatrix}
{X}_1^2 \ell_{11} + {X}_1 {X}_2 \ell_{12} + \dots + {X}_1 {X}_n \ell_{1n} \\
\vdots \\
{X}_n {X}_1 \ell_{n1} + {X}_n {X}_2 \ell_{n2} + \dots + {X}_n^2 \ell_{nn}
\end{bmatrix}, \\
(\mathbf{L} \boldsymbol{X})^2 &=
\begin{bmatrix}
{X}_1^2 \ell_{11}^2 + \dots + {X}_n^2 \ell_{1n}^2 + 2 (\ell_{11} \ell_{12} {X}_1 {X}_2 + \dots) \\
\vdots \\
{X}_1^2 \ell_{n1}^2 + \dots + {X}_n^2 \ell_{nn}^2 + 2 (\ell_{n1} \ell_{n2} {X}_1 {X}_2 + \dots)
\end{bmatrix}.
\end{align}
Given the cross-sectional independence of idiosyncratic shocks ($\mathbb{E}[X_i X_j] = 0$ for $i \neq j$) and invoking the asymptotic conditional moment bounds established in Appendix~\ref{A_A} ($\mathbb{E}[X_i \mid \boldsymbol{X} > \mathbf{b}] = o(\tau^2)$, $\mathbb{E}[X_i^2 \mid \boldsymbol{X} > \mathbf{b}] = \tau^2 + o(\tau^2)$, and $\mathbb{E}[R_3(\boldsymbol{X}) \mid \boldsymbol{X} > \mathbf{b}] = o(\tau^2)$), the conditional expectations simplify to:
\begin{align}
\mathbb{E} \left[ \langle \boldsymbol{X} \rangle \mathbf{L} \boldsymbol{X} \mid \boldsymbol{X} > \mathbf{b} \right] = \tau^2 \widehat{\mathbf{L}} + o(\tau^2),
&&
\mathbb{E} \left[ (\mathbf{L} \boldsymbol{X})^2 \mid \boldsymbol{X} > \mathbf{b} \right] = \tau^2 \widehat{\mathbf{L}\mathbf{L}^{\intercal}} + o(\tau^2),
\end{align}
where the hat operator $\widehat{\mathbf{M}}$ denotes the column vector formed by the diagonal elements of a square matrix $\mathbf{M}$, and $\mathbf{b} < \mathbf{0}$ denotes the lower threshold vector required to preserve the \textit{nonsingularity condition}.\footnote{Specifically, the multisectoral nonsingularity condition requires $\left[ \langle e^{(1-\sigma)\boldsymbol{X}}\rangle - \mathbf{A} \right]^{-1} > \mathbf{0}$. See \citet{NakanoNishimura2026} for a complete spectral characterization of this condition.}

Recalling that $\mathscr{D}(\log \mathbf{z} \mid \sigma) = -\boldsymbol{\mu}\log \mathbf{p} = -\frac{1}{1-\sigma}\boldsymbol{\mu}\ln \mathbf{p}^{1-\sigma}$, taking conditional expectations of \eqref{eq:ln_pi_gamma} yields the second-order expected DAF in a multisectoral network:
\begin{align}
\mathbb{E}\left[ \mathscr{D}(\log \mathbf{z} \mid \sigma) \mid \log \mathbf{z} > \mathbf{b} \right]
= - \boldsymbol{\mu}
\underbrace{
\left(
\mathbf{L} \widehat{\mathbf{L}} - \frac{1}{2} \mathbf{L} \mathbf{1} - \frac{1}{2} \widehat{\mathbf{L}\mathbf{L}^{\intercal}} 
\right)
}_{= \mathscr{C}}
(1-\sigma) \tau^2 + o(\tau^2).
\label{eq:expected_productivity}
\end{align}
Here, the $n$-dimensional column vector $\mathscr{C} = (\mathscr{C}_1, \dots, \mathscr{C}_n)^{\intercal}$ generalizes the scalar network \textit{impedance} coefficient of Section~2.2 to an arbitrary input-output topology. Because the production network is productive and irreducible, standard properties of the Leontief inverse ensure that $\ell_{jj} > 1$ and $\ell_{jj} \geq \ell_{ij}$ for all $i, j \in \{1, \dots, n\}$. Consequently, every sector-specific impedance component $\mathscr{C}_i$ is strictly positive:
\begin{align}
\mathscr{C}_i = \sum_{j=1}^n \ell_{ij} \ell_{jj} - \frac{1}{2}\ell_{ij} - \frac{1}{2}\ell_{ij}^2
=\frac{1}{2}\sum_{j=1}^n \ell_{ij} \left( (\ell_{jj} -1) + (\ell_{jj} - \ell_{ij}) \right) > 0.
\end{align}
Since $\boldsymbol{\mu} > \mathbf{0}$, Equation \eqref{eq:expected_productivity} establishes a fundamental structural isomorphism with Jensen's inequality in expected utility theory: an inelastic production network ($\sigma < 1$) behaves as risk-averse, strictly depressing expected aggregate productivity under symmetric microeconomic volatility (\textit{mean sinking}), whereas a Cobb-Douglas network ($\sigma = 1$) remains risk-neutral with zero mean bias.

\subsection*{Network Primarity, Spectral Scaling, and Impedance Sensitivity}

To evaluate how the depth of roundabout supply chains affects multisectoral network impedance, we generalize the notion of network \textit{primarity} (the degree to which production relies directly on primary factors rather than intermediate inputs). Define a normalized non-negative square matrix $\mathbf{P}$ as:
\begin{align}
\mathbf{P} = \langle \mathbf{1} - \mathbf{v} \rangle^{-1} \mathbf{A}.
\end{align}
Since $\mathbf{A} \mathbf{1} + \mathbf{v} = \mathbf{1}$, it follows immediately that $\mathbf{P} \mathbf{1} = \mathbf{1}$, so $\mathbf{P}$ is row-stochastic. Because $\mathbf{P}$ is the product of an irreducible non-negative matrix $\mathbf{A}$ and a strictly positive diagonal matrix $\langle \mathbf{1} - \mathbf{v} \rangle^{-1}$, $\mathbf{P}$ is itself irreducible. By the Perron-Frobenius theorem, its dominant eigenvalue is unity, with associated left and right Perron vectors satisfying $\boldsymbol{\varpi}^{\intercal} = \boldsymbol{\varpi}^{\intercal} \mathbf{P}$ and $\mathbf{1} = \mathbf{P} \mathbf{1}$, normalized such that $\boldsymbol{\varpi}^{\intercal} \mathbf{1} = 1$.

We parameterize the input-output coefficient matrix by the scalar primarity index $\kappa$ as:
\begin{align}
\mathbf{A}(\kappa) = \langle \mathbf{1} - \kappa \mathbf{v} \rangle \mathbf{P}. \label{eq:A_kappa}
\end{align}
For any $\kappa \geq 0$ satisfying $\mathbf{1} - \kappa \mathbf{v} \geq \mathbf{0}$, $\mathbf{A}(\kappa)$ remains non-negative and irreducible, and setting $\kappa = 1$ recovers the empirical status quo network $\mathbf{A}(1) = \mathbf{A}$. Decreasing $\kappa$ reduces direct primary-factor shares across all sectors, thereby lengthening roundabout intermediate production chains (increasing \textit{circularity} or outsourcing).

To examine the marginal impact of altering supply-chain depth around $\kappa = 1$, let $\rho(\kappa)$ denote the spectral radius (Perron root) of $\mathbf{A}(\kappa)$, with corresponding right and left Perron eigenvectors $\mathbf{r}(\kappa)$ and $\mathbf{l}(\kappa)^{\intercal}$ satisfying $\mathbf{A}(\kappa) \mathbf{r}(\kappa) = \rho(\kappa) \mathbf{r}(\kappa)$ and $\mathbf{l}(\kappa)^{\intercal} \mathbf{A}(\kappa) = \rho(\kappa) \mathbf{l}(\kappa)^{\intercal}$, normalized so that $\mathbf{l}(\kappa)^{\intercal} \mathbf{r}(\kappa) = 1$. Differentiating the right eigenvalue relation with respect to $\kappa$ and evaluating at $\kappa = 1$ gives:
\begin{align}
\mathbf{A}'(1) \mathbf{r}(1) + \mathbf{A}(1) \mathbf{r}'(1) = \rho'(1) \mathbf{r}(1) + \rho(1) \mathbf{r}'(1).
\end{align}
Pre-multiplying both sides by $\mathbf{l}(1)^{\intercal}$ and utilizing $\mathbf{l}(1)^{\intercal} \mathbf{A}(1) = \rho(1) \mathbf{l}(1)^{\intercal}$ together with $\mathbf{l}(1)^{\intercal} \mathbf{r}(1) = 1$ yields:
\begin{align}
\mathbf{l}(1)^{\intercal} \rho'(1) \mathbf{r}(1) &= \mathbf{l}(1)^{\intercal} \left( \mathbf{A}'(1) \mathbf{r}(1) + \mathbf{A}(1) \mathbf{r}'(1) - \rho(1) \mathbf{r}'(1) \right), \\
\rho'(1) &= \mathbf{l}^{\intercal} \mathbf{A}'(1) \mathbf{r} = - \mathbf{l}^{\intercal} \langle \mathbf{v} \rangle \mathbf{P} \mathbf{r} \equiv - {\varDelta}, \label{eq:rho_prime}
\end{align}
where $\mathbf{r} = \mathbf{r}(1)$, $\mathbf{l} = \mathbf{l}(1)$, and $\mathbf{A}'(\kappa) = - \langle \mathbf{v} \rangle \mathbf{P}$. Because $\mathbf{l}, \mathbf{r} > \mathbf{0}$, $\mathbf{v} > \mathbf{0}$, and $\mathbf{P}$ is non-negative and irreducible, the constant ${\varDelta}$ is strictly positive. A first-order Taylor expansion of the Perron root around $\kappa = 1$ therefore yields $\rho(\kappa) \approx \rho - (\kappa - 1){\varDelta}$, or equivalently:
\begin{align}
1 - \rho(\kappa) \approx 1 - \rho - (1 - \kappa){\varDelta}.
\end{align}

To project this spectral shift onto the parameterized Leontief inverse $\mathbf{L}(\kappa) = [\mathbf{I} - \mathbf{A}(\kappa)]^{-1}$, we introduce a mild spectral-gap assumption: around the status quo $\kappa = 1$, the comparative-static scaling of $\mathbf{L}(\kappa)$ is dominated by its Perron projection $(1 - \rho(\kappa))^{-1}$. Under this dominant-eigenvalue approximation, the perturbed Leontief inverse scales proportionally to the baseline matrix $\mathbf{L} = \mathbf{L}(1)$:
\begin{align}
\mathbf{L}(\kappa) \approx \mathbf{L} \frac{1 - \rho}{1 - \rho(\kappa)} 
= \mathbf{L} \frac{1 - \rho}{1 - \rho - (1 - \kappa){\varDelta}}. \label{eq:L_kappa_approx}
\end{align}
Substituting \eqref{eq:L_kappa_approx} into the impedance vector $\mathscr{C}(\kappa) = \mathbf{L}(\kappa)\widehat{\mathbf{L}(\kappa)} - \frac{1}{2}\mathbf{L}(\kappa)\mathbf{1} - \frac{1}{2}\widehat{\mathbf{L}(\kappa)\mathbf{L}(\kappa)^{\intercal}}$ yields its functional approximation in the neighborhood of $\kappa = 1$:
\begin{align}
\mathscr{C}(\kappa) 
\approx 
\left( 
\mathbf{L} \widehat{\mathbf{L}} - \frac{1}{2}\widehat{\mathbf{L}\mathbf{L}^{\intercal}} 
\right) 
\frac{(1 - \rho)^2}{(1 - \rho - (1 - \kappa){\varDelta})^2} 
- \frac{1}{2} \mathbf{L} \mathbf{1} 
\frac{1 - \rho}{1 - \rho - (1 - \kappa){\varDelta}}. 
\label{eq:C_kappa_around_1}
\end{align}

To determine whether deepening the supply chain amplifies or dampens the macroeconomic mean-sinking penalty, we differentiate \eqref{eq:C_kappa_around_1} with respect to $\kappa$ and evaluate the derivative at the status quo $\kappa = 1$:
\begin{align}
\left. \frac{\partial \mathscr{C}}{\partial \kappa} \right|_{\kappa=1} 
=
\mathscr{C}^\prime(1)
=
- \frac{\varDelta}{2(1-\rho)}
\left(
4 \mathbf{L} \widehat{\mathbf{L}} - 2 \widehat{\mathbf{L}\mathbf{L}^{\intercal}}
- \mathbf{L} \mathbf{1} 
\right).
\end{align}
Expanding the $i$th component of the bracketed vector element by element gives:
\begin{align}
\left( 4 \mathbf{L} \widehat{\mathbf{L}} - 2 \widehat{ \mathbf{L} \mathbf{L}^{\intercal} } - \mathbf{L} \mathbf{1} \right)_i
= \sum_{j=1}^n \left( 4 \ell_{jj}\ell_{ij} - 2\ell_{ij}^2 - \ell_{ij} \right) 
= \sum_{j=1}^n \ell_{ij} \left( 2 \left( 2 \ell_{jj} - \ell_{ij} \right) - 1 \right) > 0,
\label{eq:bracket_expansion}
\end{align}
where strict positivity follows from $\ell_{jj} > 1$ and $\ell_{jj} \geq \ell_{ij}$ for all $i, j$. Since $\varDelta > 0$ and $\rho \in (0, 1)$, we conclude that every component of $\mathscr{C}'(1)$ is strictly negative:
\begin{align}
\mathscr{C}^\prime(1) < \mathbf{0}.
\end{align}
Thus, increasing network primarity $\kappa$ strictly reduces multisectoral network impedance—or conversely, lengthening intermediate supply chains (increasing circularity) strictly magnifies the aggregate productivity loss under input complementarity.

\subsection*{The Cobb-Douglas Benchmark under Varying Primarity}

Finally, we contrast the CES results above with the frictionless Cobb-Douglas benchmark ($\sigma = 1$) under parameterized primarity $\kappa$. In this log-linear limiting case, the Domar Aggregation Function reduces to:
\begin{align}
\mathscr{D}(\log \mathbf{z} \mid 1 , \kappa) 
= - \boldsymbol{\mu} \mathbf{L}(\kappa) \log \mathbf{z}
= - \boldsymbol{\lambda}(\kappa)^{\intercal} \log \mathbf{z},
\end{align}
where $\boldsymbol{\lambda}(\kappa)^{\intercal} \equiv \boldsymbol{\mu} \mathbf{L}(\kappa)$ denotes the row vector of parameterized Domar weights. Suppose sector-specific log-productivity shocks follow independent zero-mean Gaussian distributions $\log {z}_i \sim \mathcal{N}(0, \tau_i^2)$ with heterogeneous volatilities $\tau_i > 0$. Because the equilibrium price mapping is globally log-linear, nonsingularity holds unconditionally across the entire support of $\log \mathbf{z}$, and the unconditional mean and variance of aggregate productivity are given by:
\begin{align}
\mathbb{E}\left[\mathscr{D}(\log \mathbf{z} \mid 1 , \kappa ) \right] = 0,
&&
\Var\left( \mathscr{D}(\log \mathbf{z} \mid 1 ,\kappa)  \right)
=
\sum_{i=1}^n (\lambda_i(\kappa) \tau_i)^2.
\end{align}

Substituting the spectral approximation \eqref{eq:L_kappa_approx} into the Domar weight vector yields:
\begin{equation}
    \boldsymbol{\lambda}(\kappa) \approx \left( \frac{1 - \rho}{1 - \rho - (1 - \kappa)\varDelta} \right) \boldsymbol{\lambda}(1).
\end{equation}
Consequently, the variance of the aggregate productivity shock at primarity $\kappa$ scales quadratically relative to its baseline variance at $\kappa = 1$:
\begin{equation}
\Var\left( \mathscr{D}(\log \mathbf{z} \mid 1 , \kappa ) \right)
\approx 
\left( \frac{1 - \rho}{1 - \rho - (1 - \kappa)\varDelta} \right)^2 
\Var\left( \mathscr{D}(\log \mathbf{z} \mid 1 , 1 ) \right).
\end{equation}
Differentiating the scaling factor with respect to $\kappa$ confirms that:
\begin{align}
\frac{\partial}{\partial \kappa}\left( \frac{1 - \rho}{1 - \rho - (1 - \kappa)\varDelta} \right)^2 
=-\frac{2(1-\rho)^2\varDelta}
{(1-\rho-(1-\kappa)\varDelta)^3} < 0.
\end{align}
In a Cobb-Douglas network, therefore, increasing primarity $\kappa$ dampens macroeconomic volatility (and deepening supply chains amplifies it), while leaving expected aggregate productivity invariant at zero. By contrast, when sectors exhibit empirical input complementarity ($\sigma < 1$), changes in primarity simultaneously govern both macroeconomic volatility and the structural mean-sinking penalty.
\section{Global Contraction of the Equilibrium Pricing Mapping} \label{A_C}

In this appendix, we prove that transforming the general equilibrium pricing system into logarithmic price coordinates restores strict contractivity in the $\ell_\infty$ norm for arbitrary real elasticities $\sigma_i \in \mathbb{R}$, thereby guaranteeing uniqueness and global convergence of fixed-point iteration whenever an equilibrium exists.

Consider the unit cost pricing equation for sector $i \in \{1, \dots, n\}$ from Equation~\eqref{production_i}:
\begin{align}
p_i = (z_i)^{-1} 
\biggl( 
\sum_{j=0}^{n} \alpha_{ij} (p_j)^{1-\sigma_i} 
\biggr)^{\frac{1}{1-\sigma_i}},
\end{align}
where $\alpha_{ij} \geq 0$ with $\sum_{j=0}^{n} \alpha_{ij} = 1$, and the primary factor is chosen as the numeraire ($p_0 = 1$). Taking natural logarithms of both sides gives
\begin{align}
\log p_i = -\log z_i + \frac{1}{1-\sigma_i} \log 
\biggl( \sum_{j=0}^{n} \alpha_{ij} e^{(1-\sigma_i)\log p_j} 
\biggr).
\end{align}
Define the vector of log sectoral prices as $\mathbf{x} = (x_1, \dots, x_n)^\intercal \in \mathbb{R}^n$ with $x_i \equiv \log p_i$ for $i = 1, \dots, n$, and set $x_0 \equiv \log p_0 = 0$. The equilibrium pricing system is then represented by the mapping $\mathbf{f} = (f_1, \dots, f_n)^\intercal : \mathbb{R}^n \to \mathbb{R}^n$, where each component function $f_i(\mathbf{x})$ is defined for $\sigma_i \neq 1$ by
\begin{align}
x_i = -\log z_i + \frac{1}{1-\sigma_i} \log 
\biggl( \sum_{j=0}^{n} \alpha_{ij} e^{(1-\sigma_i)x_j} 
\biggr) \equiv f_i(\mathbf{x}),
\end{align}
and in the Cobb-Douglas limit ($\sigma_i = 1$) by $f_i(\mathbf{x}) = -\log z_i + \sum_{j=0}^n \alpha_{ij} x_j$.

Differentiating $f_i(\mathbf{x})$ with respect to $x_j$ ($j = 0, 1, \dots, n$), the factor $(1-\sigma_i)$ cancels identically regardless of the sign of $\sigma_i$. By Shephard's lemma, each partial derivative equals the endogenous cost share $s_{ij}(\mathbf{x})$ of input $j$ in sector $i$'s total cost:
\begin{align}
\frac{\partial f_i(\mathbf{x})}{\partial x_j} 
= \frac{\alpha_{ij} e^{(1-\sigma_i)x_j}}{\sum_{k=0}^{n} \alpha_{ik} e^{(1-\sigma_i)x_k}} 
= \frac{\partial \log p_i}{\partial \log p_j} 
= s_{ij}(\mathbf{x}) \geq 0.
\end{align}
Because cost shares sum to unity across all inputs including the primary factor, it holds for every $\mathbf{x} \in \mathbb{R}^n$ and every sector $i = 1, \dots, n$ that
\begin{align}
\frac{\partial f_i(\mathbf{x})}{\partial x_0} + \sum_{j=1}^{n} \frac{\partial f_i(\mathbf{x})}{\partial x_j} = 1.
\label{eq:cost_share_sum}
\end{align}
Whenever every sector uses a strictly positive primary-factor share ($\alpha_{i0} > 0$), we have $\partial f_i(\mathbf{x})/\partial x_0 = s_{i0}(\mathbf{x}) > 0$ for all finite $\mathbf{x} \in \mathbb{R}^n$, which implies $\sum_{j=1}^{n} \partial f_i(\mathbf{x})/\partial x_j = 1 - s_{i0}(\mathbf{x}) < 1$.

Now let $\mathbf{u}, \mathbf{w} \in \mathbb{R}^n$ be any two distinct log-price vectors ($\mathbf{u} \neq \mathbf{w}$), and parameterize the line segment connecting them as
\begin{align}
\mathbf{x}(t) = \mathbf{u} + t(\mathbf{w} - \mathbf{u}), \qquad t \in [0, 1].
\end{align}
For each sector $i \in \{1, \dots, n\}$, define the scalar composite function $F_i(t) \equiv f_i(\mathbf{u} + t(\mathbf{w} - \mathbf{u}))$. Since $F_i$ is continuously differentiable on $[0, 1]$, the Mean Value Theorem guarantees the existence of some $t_i \in (0, 1)$ such that
\begin{align}
F_i(1) - F_i(0) = \frac{d F_i(t_i)}{dt}.
\end{align}
Applying the multivariate chain rule to the right-hand side yields
\begin{align}
f_i(\mathbf{w}) - f_i(\mathbf{u}) 
= \sum_{j=1}^{n} \frac{\partial f_i(\mathbf{x}(t_i))}{\partial x_j} (w_j - u_j).
\end{align}
Since $\partial f_i(\mathbf{x}(t_i))/\partial x_j = s_{ij}(\mathbf{x}(t_i)) \geq 0$ for all $j = 1, \dots, n$ and $\sum_{j=1}^{n} \partial f_i(\mathbf{x}(t_i))/\partial x_j = 1 - s_{i0}(\mathbf{x}(t_i)) < 1$ by Equation~\eqref{eq:cost_share_sum}, taking absolute values and bounding $|w_j - u_j| \leq \|\mathbf{w} - \mathbf{u}\|_\infty \equiv \max_{1 \leq k \leq n} |w_k - u_k|$ gives
\begin{align}
|f_i(\mathbf{w}) - f_i(\mathbf{u})| 
\leq 
\sum_{j=1}^{n} \frac{\partial f_i(\mathbf{x}(t_i))}{\partial x_j} |w_j - u_j| 
\leq 
\sum_{j=1}^{n} \frac{\partial f_i(\mathbf{x}(t_i))}{\partial x_j} 
\|\mathbf{w} - \mathbf{u}\|_\infty
< \|\mathbf{w} - \mathbf{u}\|_\infty.
\end{align}
Because this strict inequality holds for every component $i = 1, \dots, n$, taking the maximum over $i$ establishes that
\begin{align}
\|\mathbf{f}(\mathbf{w}) - \mathbf{f}(\mathbf{u})\|_\infty < \|\mathbf{w} - \mathbf{u}\|_\infty \qquad \forall \, \mathbf{u} \neq \mathbf{w} \in \mathbb{R}^n.
\label{eq:strict_contraction}
\end{align}

Equation~\eqref{eq:strict_contraction} proves that $\mathbf{f}$ is a \textit{strict contraction} 
with respect to the $\ell_\infty$ norm across the entire log-price space $\mathbb{R}^n$, irrespective of the signs of the sectoral substitution elasticities $\boldsymbol{\sigma} = (\sigma_1, \dots, \sigma_n)$. Consequently, if an equilibrium log-price vector $\mathbf{x}^* = \mathbf{f}(\mathbf{x}^*)$ exists, it is necessarily unique. Furthermore, on any compact convex subset $\mathcal{K} \subset \mathbb{R}^n$ containing the equilibrium trajectory, the contraction modulus $\bar{\lambda}_{\mathcal{K}} \equiv \max_{1 \leq i \leq n} \max_{\mathbf{x} \in \mathcal{K}} \bigl(1 - s_{i0}(\mathbf{x})\bigr)$ is strictly less than unity; by the Banach fixed-point theorem, successive approximations $\mathbf{x}^{(k+1)} = \mathbf{f}(\mathbf{x}^{(k)})$ converge globally and geometrically to the unique general equilibrium price vector $\mathbf{p}^* = \exp(\mathbf{x}^*)$ from any initial guess.


\bibliographystyle{apalike}
\raggedright
\bibliography{BIBFILE}

@article{Acemoglu2017,
  author    = {Acemoglu, Daron and Ozdaglar, Asuman and Tahbaz-Salehi, Alireza},
  title     = {Microeconomic Origins of Macroeconomic Tail Risks},
  journal   = {American Economic Review},
  volume    = {107},
  number    = {1},
  pages     = {54--108},
  year      = {2017}
}

@article{BaqaeeFarhi2019,
  author    = {Baqaee, David Rezza and Farhi, Emmanuel},
  title     = {The Macroeconomic Impact of Microeconomic Shocks: Beyond {Hulten}'s Theorem},
  journal   = {Econometrica},
  volume    = {87},
  number    = {4},
  pages     = {1155--1203},
  year      = {2019}
}

@book{Krasnoselskii1964,
  author    = {Krasnosel'ski{\u{\i}}, Mark Aleksandrovich},
  title     = {Positive Solutions of Operator Equations},
  publisher = {P. Noordhoff},
  address   = {Groningen},
  year      = {1964}
}

@book{Lucas1987,
  author    = {Lucas, Robert E.},
  title     = {Models of Business Cycles},
  publisher = {Basil Blackwell},
  address   = {Oxford},
  year      = {1987}
}

@article{NakanoNishimura2026,
  author    = {Nakano, Satoshi and Nishimura, Kazuhiko},
  title     = {Nonlinear {Domar} Aggregation over Transforming Production Networks},
  journal   = {Journal of Evolutionary Economics},
  volume    = {36},
  number    = {2},
  pages     = {58},
  year      = {2026}
}

\end{document}